\documentclass[12pt]{article}  
\usepackage{amsmath, amssymb, amsthm}  
\usepackage{graphicx}  
\usepackage{booktabs}  
\usepackage{float}  
\usepackage[margin=1in]{geometry}  
\usepackage{setspace}  
\usepackage{natbib}  
\usepackage{hyperref}  
\usepackage{enumitem}  
\usepackage{caption, subcaption}  
\usepackage{lmodern}  
\usepackage{mathtools}  
\usepackage{dsfont}  
\usepackage{xcolor}  
\usepackage{bm}  
\usepackage{algorithm}
\usepackage{algpseudocode}
\usepackage{authblk} 
\usepackage{float}
\usepackage{comment}
\usepackage{adjustbox} 
\usepackage{lscape}    
\usepackage{longtable} 
\usepackage{pdflscape} 
\usepackage{rotating}
\usepackage{placeins}  
\usepackage{multirow}  
\usepackage{algorithm}
\usepackage{algpseudocode}

\theoremstyle{plain}

\theoremstyle{definition}

\theoremstyle{remark}

\newcommand{\independent}{\perp\!\!\!\perp}

\hypersetup{
    colorlinks=true,
    linkcolor=blue,
    citecolor=blue,
    urlcolor=blue
}

\usepackage{tcolorbox}
\tcbset{colback=white, colframe=black, boxrule=0.5pt, arc=2pt, left=4pt, right=4pt, top=4pt, bottom=4pt}
\setlist{nosep}

\begin{document}

\title{A New Perspective of the Meese-Rogoff Puzzle: Application of Sparse Dynamic Shrinkage\thanks{This research was supported by The University of Melbourne’s Research Computing Services and the Petascale Campus Initiative.}}

\author[1]{Zheng Fan\thanks{Department of Economics, University of Melbourne. \\ 
    Email: \href{mailto:zhengf1@student.unimelb.edu.au}{zhengf1@student.unimelb.edu.au}.
    }}
\author[2]{Worapree Maneesoonthorn\thanks{Maneesoonthorn is supported by the ARC Discovery Grant DP200101414.}}
\author[1]{Yong Song\thanks{Yong Song thanks for the support from ARC Discovery Grant DP230100959.}}

\affil[1]{\small Department of Economics, University of Melbourne}
\affil[2]{\small Department of Econometrics and Business Statistics, Monash University}

\vspace{-0.5cm}
\date{\today}
\maketitle
\vspace{-1cm}
\begin{abstract}
\begingroup
\setstretch{1.2} 

We propose the Markov Switching Dynamic Shrinkage process (MSDSP), nesting the Dynamic Shrinkage Process (DSP) of \cite{kowal2019dynamic}. We revisit the Meese-Rogoff puzzle 
\citep{meese1983empirical,meese1983out,meese1988was} by applying the MSDSP to the economic models deemed inferior to the random walk model for exchange rate predictions. The flexibility of the MSDSP model captures the possibility of zero coefficients (sparsity), constant coefficient (dynamic shrinkage), as well as sudden and gradual parameter movements (structural change) in the time-varying parameter model setting. We also apply MSDSP in the context of Bayesian predictive synthesis (BPS)  \citep{mcalinn2019dynamic}, where dynamic combination schemes exploit the information from the alternative economic models. Our analysis provide a new perspective to the Meese-Rogoff puzzle, illustrating that the economic models, enhanced with the parameter flexibility of the MSDSP, produce predictive distributions that are superior to the random walk model, even when stochastic volatility is considered. 

\endgroup
\end{abstract}



\section{Introduction}
Forecasting exchange rate is important for policy makers, investors and international
traders. The seminal Meese–Rogoff Puzzle of
\cite{meese1983empirical} claims that no model can systematically beat the random walk, with subsequent literature predominantly focusing on the point forecast. Thirty years on, \cite{rossi2013exchange} confirmed such claim
with an exhaustive empirical work. 
On the other hand, \cite{rossi2013exchange} also inspires that
\begin{quote}
    The efficient market hypothesis does not mean that exchange rates are unrelated
to economic fundamentals, nor that exchange rates should fluctuate randomly
around their past values.
\end{quote}
This work motivates economic theoretic driven studies such as \cite{ferraro2015can,cheung2019exchange,engel2019uncovered,candian2023imperfect,neghab2024explaining} amongst others. A more recent and comprehensive survey is conducted in \cite{fang202430}, who thoroughly reviewed the literature in this area. They found no influential findings after \cite{rossi2013exchange},
with the exception of \cite{ferraro2015can},
even though many researchers have attempted to address the puzzle.

Challenges in exchange rate forecasting include: 
(1) dynamic instability, (2) limited sample size, and (3) model uncertainty.
There are other concerns such as data frequency choice 
and the discussion of big and small economy, 
but we focus on the aforementioned three points in this paper.
Our novel econometric approach  addresses these three challenges 
and indeed provide a dominating predictive outcome in 
many out-of-sample metrics than the random walk.
The foundation of our econometric models is built upon established economic theories, with historical research on the relationship between exchange rates and economic fundamentals respected in our framework.

From the econometric modelling perspective, many research addressed dynamic instability. Existing popular method such as time-varying parameter (TVP) models 
\cite{wolff1987time,canova1993modelling,mumtaz2013time,byrne2016exchange}
cater for gradual change of the parameters
and regime-switching models \cite{engel1994can,nikolsko2012markov}
allow for sudden parameter change.
These models are flexible and well-tested in time series forecasting.
However, to our knowledge, 
no paper has seriously addressed the importance of model parsimony 
in the presence of limited sample size for exchange rate forecasting.
A data-driven balance between flexibility and parsimony, in our view,
could shed lights on the Meese–Rogoff prediction puzzle.

To answer challenges (1) and (2),
we consider the dynamic shrinkage process (DSP) of \cite{kowal2019dynamic}.
The DSP is a process taking the form of a state space model,
with the innovation taking a particular form such that
the process could experience a decent duration of large negative values. See also
\cite{hauzenberger2024dynamic}, who applied DSP to their global
shrinkage factor. When used to model the log volatility of a dynamic parameter, the DSP allows for the variation of the time-varying parameter to approach zero, reducing its temporal variation. This approach shrinks the parameters towards the value in the previous period rather than shrinking them to zero.
The same idea can be found in \cite{dufays2021sparse},
but their model is computationally intensive because it requires a joint movements of the parameters at adjacent time points.
In contrast, the DSP works as a clip by collecting adjacent
parameters into a single value through the innovation's volatility, which is a stark contrast to the standard TVP model that assumes constant volatility.
Another closely related research is \cite{knaus2023dynamic},
who proposed dynamic triple gamma prior to model innovations in the dynamic linear model framework.

In addition to shrinkage of volatility, \cite{huber2021inducing} emphasized 
the empirical importance of sparsity. They pointed out that
setting irrelevant parameters to zero may greatly improve prediction.
Motivated by \cite{west2006bayesian},
\cite{kalli2014time} developed a Normal Gamma Autoregressive process (NGAR)
that can mask segments of a time series of coefficients to be numeric zero.
Similar ideas include the dynamic slab and spike process in \cite{rockova2021dynamic}
and the Markov mask of latent coefficients in
\cite{uribe2020dynamic} and \cite{bernardi2023dynamic}.\footnote{
The idea of sparsity via masking can be traced back to \cite{nakajima2013bayesian}.}
\cite{lopes2022parsimony}, instead, use a 4-component mixture prior to select between TVP and constant coefficient models, but they assume static mixtures. There is no dynamic framework to date that allow both sparsity and (potentially) multiple regimes of non-zero, but constant, coefficient in the TVP setting. 

\begin{figure}
    \centering
    \includegraphics[width=1\linewidth]{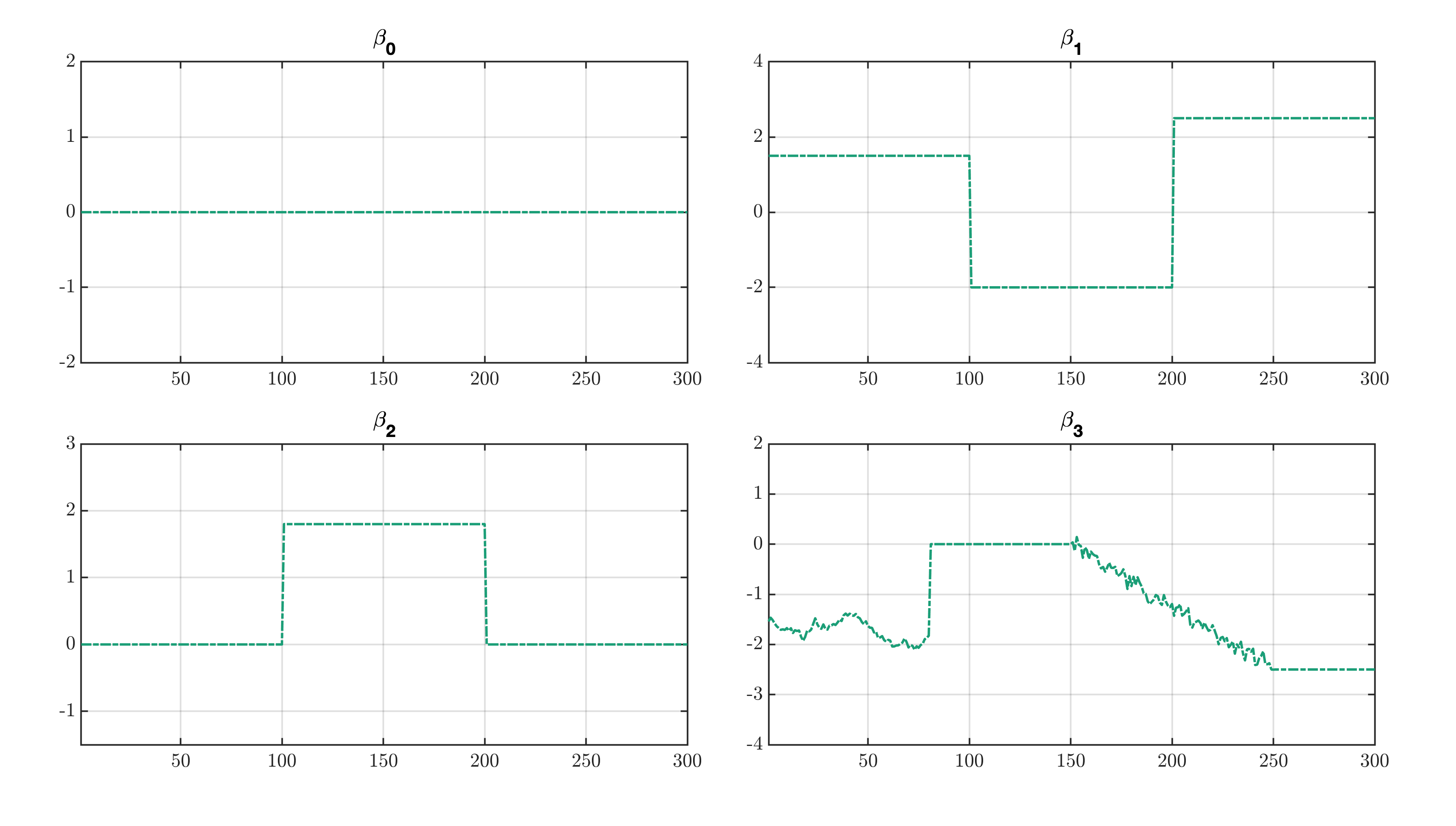}
    \caption{Four synthetic parameter paths}
    \label{fig:sim_true}
\end{figure}

We bridge this gap by introducing the Markov Switching DSP (\textbf{MSDSP}), which allows for the time-varying coefficient to dynamically switch between the zero-state (achieving sparsity) and the DSP state (achieving shrinkage).  To illustrate the flexibility of our proposed framework, we plot four synthetic parameter paths in Figure~\ref{fig:sim_true}.
The first parameter $\beta_0$ is a constant $0$ which renders its corresponding covariate (or intercept) useless.
The second parameter $\beta_1$ changes values twice, with all three values being none-zero. The third parameter $\beta_2$ is zero at the two ends of the illustrative period, and is only ``activated'' in the middle period. The last parameter $\beta_3$ experiences gradual change, becomes abruptly ``inactivated'' at zero,
and then gradually move towards a fixed none-zero value.
All these behaviors can be captured by our MSDSP framework.

In the exchange rate forecasting application, we follow the established benchmark models from \cite{rossi2013exchange} and find that when we apply our proposed MSDSP method to the economic models, they substantially outperform the random walk across a range of metrics. In addition applying the MSDSP to enhance the individual models, we find that the MSDSP naturally lends itself to the  Bayesian predictive synthesis (BPS).
BPS is a novel and powerful model assembly method developed by \cite{mcalinn2019dynamic}.  It takes care of potential model bias and 
allows unrestricted model weights (even negative) to optimally combine
many sub-models or experts, see \cite{mcalinn2020multivariate,tallman2024bayesian} for recent applications. In the BPS context, the MSDSP can be used directly on
model weights in model assembly. 
These weights can move as in a state space model with parsimony (DSP-state)
or simply shut down (zero-state) for a while and then reactivate (or not).
We apply the MSDSP assembly method and compare it to 
some existing approaches. We find
that the MSDSP assembly method works the best 
among multiple well-known assembly methods.
With this, we answer to the aforementioned challenge (3).

The remainder of the paper is organized as follows. Section 2 outlines the construction of the MSDSP framework for TVP models, along with the Bayesian inference proposed for the the framework. A simulation study to assess the advocacy of the framework is provided in Section 3. In Section 4, focus is given to a detailed application of the MSDSP framework on existing economic models to address the Meese-Rogoff puzzle, while Section 5 provides an application of the MSDSP in model assembly for exchange rate forecasting. Section 6 concludes.

\section{The Markov Switching Dynamic Shrinkage Process}
\label{sec:model}
The Markov Switching Dynamic Shrinkage Process (MSDSP) provides enhanced flexibility to the state space model that encapsulates the TVP specificaiton in many macroeconomic models. The state-space model, tracing back to \cite{kalman1960new}, is a statistical framework 
for dynamic systems, with the most commonly used being the Gaussian Dynamic Linear Model (DLM) \citep{harrison1976bayesian, west2006bayesian}, assuming linear relationship between the observables and the state, with normally distributed innovations. The TVP model for time series analysis belongs to this class, and can be specified as:
\begin{align}
    y_t &= \bm{x}'_t\bm{\beta}_t +\epsilon_t, & \epsilon_t \sim N\left(0,\sigma_t^2 \right), \label{eq:dlm_y}\\
    \bm{\beta}_t &= \bm{\beta}_{t-1}+\bm{\omega}_t, & \bm{\omega}_t \sim N\left(0, \Omega_t\right), \label{eq:dlm_beta}
\end{align}
where the innovation $\epsilon_t$'s variance $\sigma^2_t$ and $\bm{\omega}_t$'s covariance matrix $\Omega_t$ are pre-determined or a constant.
The variable of interest $y_t$ is a scalar, vector $\bm{x}_t$ is the $p\times 1$ observed data including the intercept.
The time-varying parameter $\bm{\beta}_t$ is a $p\times 1$ vector.

Contemporary macroeconomic or financial applications emphasized the 
time varying volatility.
For example, stochastic volatility (SV) has been advocated and evidenced
by numerous empirical works \citep{primiceri2005time, justiniano2008time, nakajima2011time,chan2018bayesian,chan2024large}.
Here, we base our model on the DLM with SV specification of  \citep{aguilar2000bayesian}, who assumes \eqref{eq:dlm_y}-\eqref{eq:dlm_beta}, but with $\sigma_t^2=\exp (g_t)$, and  
\begin{align}
    g_t &= \mu_g+\phi_g (g_{t-1}-\mu_g)+e_t, & e_t \sim N\left(0,\sigma_g^2\right),
    \label{eq:dlm_sv_g}
\end{align}
The log volatility $g_t$ follows an AR(1) process, with stationarity imposed by $|\phi_g| <1$. 
In addition, we assume a diagonal matrix for $\Omega_t$ in \eqref{eq:dlm_beta}  for simplicity.

\subsection{TVP with MSDSP Flexibility}\label{subsec:msdsp}

Our goal is to extend the TVP modelling framework to account for both sparsity and shrinkage of the coefficient to a non-zero value. To this end, we extend the work of \cite{kowal2019dynamic}, and introduce $p$ two-state Markov switching processes to allow for the time-varying coefficients to ``switch off'' to zero. The model, in full, takes the form of
\begin{align}
    y_t &= \bm{x}'_t\bm{\beta}_t +\epsilon_t, & \epsilon_t \sim N\left(0,\exp(g_t)\right) \label{eq:msdsp_y} \\
    g_t &= \mu_g+\phi_g (g_{t-1}-\mu_g)+e_t, & e_t \sim N\left(0,\sigma_g^2\right) \label{eq:msdsp_g}\\
     \bm{\beta}_{t} &= \bm{s}_t\circ \tilde{\bm{\beta}}_t + (\bm{\ell}-\bm{s}_t)\circ {\bm{0}}\label{eq:msdsp_beta}\\
     \tilde{{\beta}}_{it} &=\tilde{{\beta}}_{i,t-1} + {\omega}_{it} & \omega_{it} \sim N\left(0, \exp(h_{it})\right) \label{eq:msdsp_beta_tilde}\\
     h_{i,t+1}& = \mu_i + \phi_i (h_{it} -\mu_i) + \eta_{i,t+1},  & \eta_{i,t+1}\sim Z(\alpha_h, \beta_h, 0, 1) \label{eq:msdsp_h}\\
     P(s_{it}=k\mid s_{i,t-1}=j) & = P^i_{jk} \label{eq:msdsp_s}
\end{align}
for $i=1,..,p$, $t=1,...,T$ and $j,k\in \{0,1\}$. The $\circ$ in Equation~\eqref{eq:msdsp_beta}
is the Hadamard (pointwise) product,
and $\bm{0}$ is a $p\times 1$ vector of zeros. The introduction of the $p\times 1$ variable, $\bm{s}_t=(s_{1t},...,s_{pt})$, allow for the elements of $\bm{\beta}_t$ to take value either the value of zero or the corresponding element of the ``shadow'' coefficient $\tilde{\bm{\beta}}_t$, based on the Markovian transition probability in \eqref{eq:msdsp_s}. We assume that Markovian switches operate independently across the $p$ dimension, such that $\bm{s}_i \independent \bm{s}_j$ for $j\neq i$ and $\bm{s}_i = (s_{i1},...,s_{iT})$.

The element of the shadow coefficient $\tilde{\bm{\beta}}_t$, $\tilde{{\beta}}_{it}$ for $i=1,\dots,p$, follows the Dynamic Shrinkage Process (DSP) of \cite{kowal2019dynamic}. The DSP allows for the latent coefficient itself to exhibit SV structure through the modelling of the volatility of each innovation term $\omega_{it}$ in \eqref{eq:msdsp_beta_tilde}-\eqref{eq:msdsp_h}. \cite{kowal2019dynamic} proposed the use of the Z-distribution as a prior in \eqref{eq:msdsp_h}, with the feature of this particular distribution allowing for extreme negative values in the log volatility term $h_{it}$. The Z-distribution, introduced by \cite{Barndorff-Nielsen1982-yc}, is a flexible class of normal mixture that allows for extreme skewness and fat-tails. In the DSP context, we employ the mixture constructed using the logistic transformation of the beta variable with parameters $\alpha_h$ and $\beta_h$, with zero location and unit scale for identification. In its extremes, \( \eta_{it} \to -\infty \) results in the log-variance \( h_{j,t} \) approaches $-\infty$. In such circumstances, the volatility of  $\tilde{{\beta}}_{it}$ approaches zero, rendering a constant coefficient environment with $\tilde{{\beta}}_{it}=\tilde{{\beta}}_{i,t-1}$ .

Our proposed MSDSP specification offers additional parsimony and flexibility to the TVP model as follows. First, each element of $\bm{\beta}_t$ is governed by independent dynamic shrinkage and Markov switching processes, allowing for each covariate to behave independent through the sample period. The general framework also allows for the potential for the TVP to switch to zero for sparsity, to shrink to a constant value, or to revolve dynamically according to the state space structure. Our specification for sparsity and shrinkage is internally coherent, as also done in  \cite{rockova2021dynamic}, which is in contrast to the post-processing approaches of  \cite{huber2021inducing,Hahn2015-an}, among many.

\subsection{Nested and Related Models}
\label{subsec:dsp}

The general MSDSP framework proposed here nests several existing models. The DSP proposed by \cite{kowal2019dynamic} is its most obvious special case, which can be achieved when the transition probability \(P^i_{jj}=1\) for both \(j \in \{0,1\}\) and the two-state Markow switching process reduces to a one-state scenario. This coincide with the original TVP with DSP specification. In the case when \( \alpha_h = \beta_h = 1/2 \), and in the absence of dynamic structure, in the specification of the Z-distribution in \eqref{eq:msdsp_h}, the model corresponds to a horse-shoe prior also discussed in \cite{kowal2019dynamic}. Of course, the original TVP model is also nested within our framework, with \( \bm{\beta}_{t} = \tilde{\bm{\beta}}_t \) and the latent time-varying parameter exhibit homoskedasticity.
Lastly, if $\tilde{\bm{\beta}}_{it}$ is a constant,
the model assumes a dynamic stochastic search variable selection (SSVS) method
for the corresponding variable $\bm{x}_i$.

A closely related framework is that of \cite{bernardi2023dynamic}, who employ
one latent parameter process and one masking process, with roles similar to $\tilde{\beta}_{it}$ and $s_{it}$, respectively, in our framework. Their latent parameter process is a simple random walk (or some joint normal distribution)
instead of DSP in this paper, so their latent model cannot afford short-term constant coefficient.
The mask parameter is determined by an inverse logit function with input from another latent normal distribution, with the normality assumptions lending itself naturally the variational inference in their paper.

\subsection{Bayesian Inference of the MSDSP Model}

We use Bayesian computation to conduct inference on the MSDSP model. Using the Polya-gamma representation for the DSP \citep{polson2013bayesian} and the mixture technique for the SV specification \citep{Omori2007-is}, we seek to establish the augmented posterior:
\begin{equation}
    p(\Psi\mid \bm{y}_{1:T},\bm{x}_{1:T}),
\end{equation}
where \(\bm{y}_{1:T}=(y_1,\dots, y_T)^\top\), \(\bm{x}_{1:T}=(\bm{x}_1,\dots,\bm{x}_T)\) with \(\bm{x}_t=\{x_{it}\}_{i=1}^p\) and the model unknowns are collected in \(\Psi\). The model unknowns include any auxiliary variables from the Polya-gamma resprentation and the SV mixture representation that is amendable to the Gibbs sampler in the Markov chain Monte Carlo (MCMC) sampling to estimate our posterior distribution. The elements of \(\Psi\), along with their respective description and dimensions are summarized in Table \ref{tab:summary}

\begin{table}[htbp]
\centering
\renewcommand{\arraystretch}{1.3}
\resizebox{\textwidth}{!}{
\begin{tabular}{lll}
\hline\hline
Notation& Description& Dimension\\
\hline
$\tilde{\bm{\beta}} = (\tilde{\bm{\beta}}_1, \ldots, \tilde{\bm{\beta}}_T)$ with $\tilde{\bm{\beta}}_t=\{\tilde{\beta}_{it}\}_{i=1}^p$& Shadow coefficient & \(T \times p\)\\
\hline
$\bm{s}= (\bm{s}_1,\dots,\bm{s}_T)$ with \(\bm{s}_t = \{s_{it}\}_{i=1}^p\)& State indicator for each variable and time point & \(T \times p \)\\
\hline
$g=(g_1,\dots,g_T)^\top$ & Log volatility of the measurement equation & \(T \times 1\)\\
\hline
$\mu_g, \phi_g, \sigma^2_g$ & Parameters of the log volatility process $g_t$ & 3\\
\hline
$P = \{P^i\}_{i=1}^p$ with $P^i \in \mathbb{R}^{2 \times 2}$ & State transition probability matrices & $2p$ \\
\hline
${H}= (\bm{h}_1,\dots,\bm{h}_T)$ with \(\bm{h}_t = \{h_{it}\}_{i=1}^p\)& DSP-related log volatility & \(T \times p \)\\
\hline
$\bm{\mu} = \{\mu_i\}_{i=1}^p$, $\bm{\phi} = \{\phi_i\}_{i=1}^p$& Parameters associated with $h_{it}$ & \(2p\)\\
\hline
$\bm{\xi}_\mu = \{\xi_{\mu,i}\}_{i=1}^p$& Auxiliary variable for sampling $\mu_i$ & \(p\)\\
\hline
$\bm{\xi} = (\bm{\xi}_1,\dots,\bm{\xi}_T)$ with \(\bm{\xi_t}=\{\xi_{it}\}_{i=1}^p\)& Auxiliary variable for sampling $h_{it}$ & \(T \times p \)\\
\hline
$\bm{r} = \{r_t\}_{t=1}^T$& Auxiliary variable for sampling $g_t$ & \(T \times 1\)\\
\hline \hline
\end{tabular}
} 
\caption{Summary of the elements of $\Psi$ from the Markov Switching Dynamic Shrinkage Process model}
\label{tab:summary}
\end{table}

Within our MCMC algorithm, we utilize the techniques to sample the components related to the DSP component outlined in \cite{kowal2019dynamic}. Standard algorithms for Markov switching and SV inference are adopted. Algorithm~\ref{alg:1} outlines the steps for our MCMC inference, with full details of the algorithm and the prior distribution given in the Appendix. In all our inference, we construct \(G=4,000\) posterior samples after $30,000$ burn-in draws, with every 5th draw retained through thinning.
Simulation consistency inference can be carried out from our $G$ this posterior samples. For instance, if the posterior expected value of 
$\bm{\beta}_t$, or simply $E(\bm{\beta}_t\mid \bm{y}_{1:T},\bm{x}_{1:T})$,
can be estimated by the posterior sample mean 
$\frac{1}{G}\sum\limits_{g=1}^G \bm{\beta}_t^{(g)}$.

\begin{algorithm}[H]
\caption{MCMC. After initialization, repeat the following steps:}
\label{alg:1}
 \begin{enumerate}[label=(\alph*)]
    \item Sample from \({\tilde{\bm{\beta}}}\mid\bm{y},\bm{X}, {\bm{s}},\bm{g}\).
    \item Sample $\bm{s} \mid \bm{y},\bm{X},\tilde{\bm{\beta}}, \Sigma_\epsilon, P$ by using FFBS method. 
    \item Update $\bm{\beta}$ from Equation~\eqref{eq:msdsp_beta}.
    \item Sample $\bm{g}\mid \bm{y},\bm{X},\bm{\beta},\bm{r},\mu_g,\phi_g,\sigma_g^2$.
    \item Sample  $\mu_g,\phi_g \mid \bm{g},\sigma^2_g$ and then $\sigma^2_g\mid, \bm{g},\mu_g,\phi_g$.
    \item Sample $P\mid \bm{s}$.
    \item Sample $H\mid \tilde{\bm{\beta}},\bm{\mu},\bm{\phi},\bm{\xi}$. 
    \item Sample $\bm{\mu}\mid H, \bm{\phi}, \bm{\xi}_\mu, \bm{\xi}$ and $\bm{\phi}\mid H, \bm{\mu}, \bm{\xi}$.
    \item Sample $\bm{\xi}_\mu \mid H, \bm{\phi}, \bm{\xi}$
    \item Sample $\bm{\xi}\mid H, \bm{\mu}, \bm{\phi}$.
    \item Sample $\bm{r} \mid \bm{y},\bm{X}, \bm{\beta},\bm{g}$.
\end{enumerate}
\end{algorithm}

\subsection{Direct $h-$step-ahead Forecasts}
\label{subsec:oos}
In the application of exchange rate forecasting, our primary focus is to construct the predictive distribution for future exchange rates. The model specified in Section \ref{subsec:msdsp} specifies a contemporaneous regression model in \eqref{eq:msdsp_y}. Without an explicit dynamic for the covariate, $x_t$, constructing predictions for future time points is not feasible. In order to construct the $h-$step-ahead predictive distribution, we propose the direct $h-$step-ahead forecasting model by adjusting the measurement equation \eqref{eq:msdsp_y} to
\begin{equation}
    y_{t+h} = \bm{x}'_t\bm{\beta}_t +\epsilon_t, \label{eq:y_th}
\end{equation}
with $\epsilon_t \sim N\left(0,\exp(g_t)\right)$ and the remainder of the model components remain as defined in \eqref{eq:msdsp_g}-\eqref{eq:msdsp_s}. 

The out-of-sample $h$-step-ahead posterior predictive distribution, integrating out the uncertainty in the model unknowns, takes the form of 
\begin{equation}
    p(y_{T+h}|I_T) = \int p(y_{T+h} \mid x_T,\bm{\beta}_T,g_T)p(\Psi \mid \bm{y}_{h+1:T}, \bm{x}_{1:T-h}) d\Psi, \label{eq:postpred}
\end{equation}
where $I_T=(\bm{y}_{1:T},\bm{x}_{1:T})$ denotes the information set up to time $T$. By construction, the predictive distribution, even for $h>1$, can be constructed using the covariate observed within the sample period. In addition, the implied TVP model structure, captured by  \(p(y_{T+h} \mid x_T,\bm{\beta}_T,g_T)\), reflects the specific dynamic relationship between the exchange rate and the $h-$lag covariate without the need to further assume the evolution of such relationship between periods $T$ and $T+h$.  

Note that when applying the model in \eqref{eq:y_th}, the posterior distribution $p(\Psi \mid \bm{y}_{h+1:T}, \bm{x}_{1:T-h})$ needed in \eqref{eq:postpred} only include the inference of the latent quantities up to time point $T-h$. That is, we obtain $G$ posterior daws of $(\bm{s}_{T-h},\tilde{\bm{\beta}}_{T-h},\bm{h}_{T-h},g_{T-h})$. In order to produce the $h-$step-ahead prediction, we simulate forward using the dynamic specification in \eqref{eq:msdsp_g}-\eqref{eq:msdsp_s} to obtain $G$ draws of $\bm{\beta}_T$ and $g_T$, upon which $p(y_{T+h} \mid x_T,\bm{\beta}_T,g_T)$ conditions on. By simulation, we obtain $G$ draws of $y_{T+h}$, 
$\{y^{(g)}_{T+h}\}_{g=1}^G$, from its conditional predictive distribution. All predictive statistics in the application
can be inferred from this framework. For example, the expected value $E(y_{T+h}\mid I_T)$
can be estimated by the sample mean $\frac{1}{G}\sum\limits_{g=1}^G y_{T+h}^{(g)}$. The predictive density $p(y_{T+h}\mid I_T)$
can be estimated by the sample mean of the Gaussian conditional densities implied 
by \eqref{eq:y_th} as $\frac{1}{G}\sum\limits_{g=1}^G p(y_{T+h}\mid x_T, \bm{\beta}^{(g)}_T, g_T^{(g)})$, evaluated over the plausible grid value of $y_{T+h}$. We evaluate the prediction using five metrics geared to evaluate the predictive distribution. These metrics are defined and discussed in Section~\ref{subsec:metrics}.

\section{Simulation Study}
\label{sec:sim}

We carried out two simulation studies to illustrate the 
effectiveness of the MSDSP framework. In particular, we highlight the ability of the MSDSP in flexibly capturing multiple regimes, with the posterior inference of these dynamic parameters efficient for both sudden and gradual shifts in the TVP. 

\subsection{Case 1: Abrupt Shifts in TVP}
\begin{figure}
    \centering
\includegraphics[width=1\textwidth]{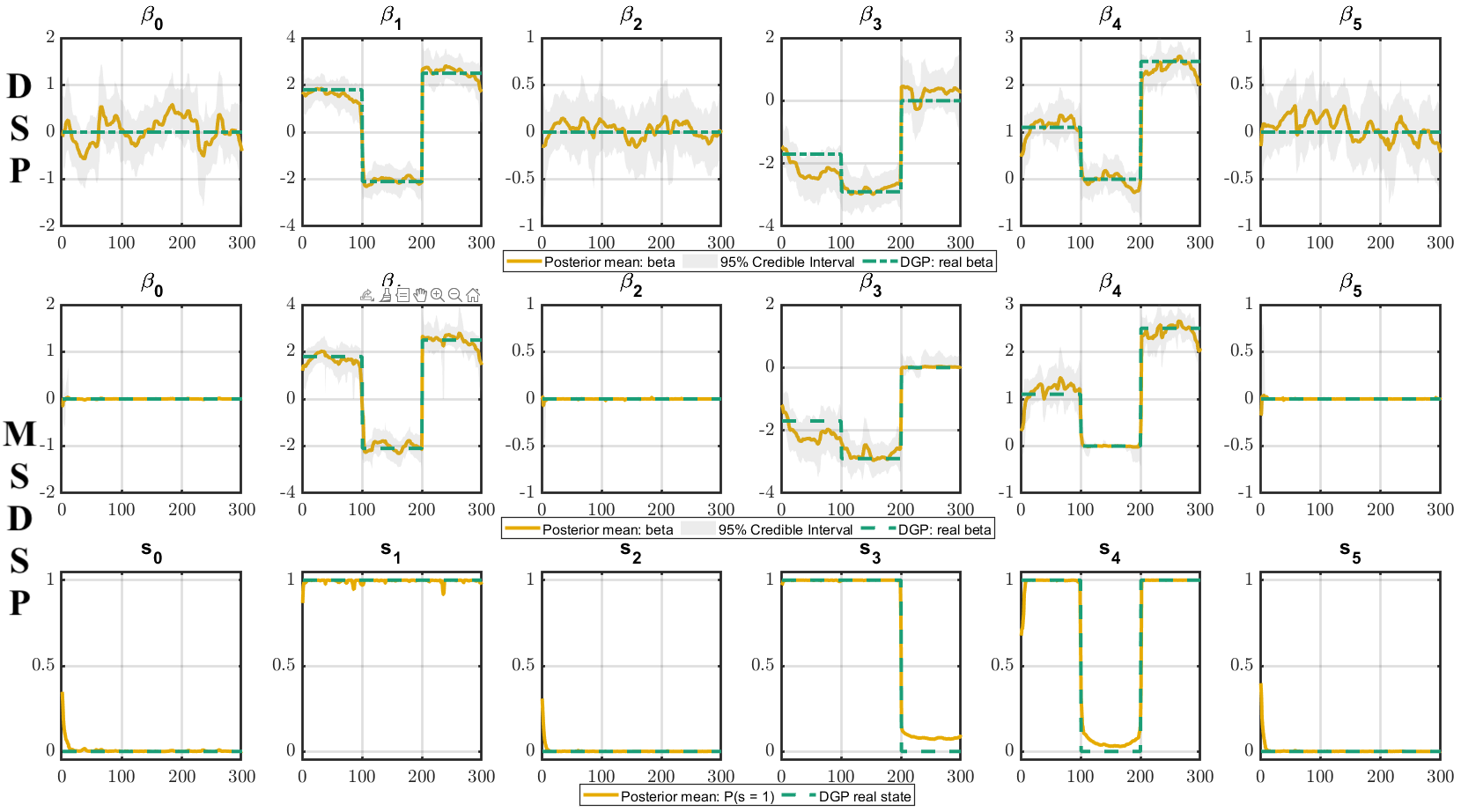}
\caption{Posterior estimates from DSP and MSDSP models on DGP 1. 
The first two rows show the posterior mean and 95\% credible intervals of 
$\bm{\beta}_t$ from DSP and MSDSP, respectively. The green dashed lines are the true values.
The bottom row shows the posterior probabilities of being in the DSP-state 
in the MSDSP model. The green dashed lines are the indicators of the true DSP-state.}
     \label{fig:simu_msdsp_beta}
\end{figure}

We consider a scenario with $5$ explanatory variables and an intercept, 
for $T = 300$ periods. 
The explanatory variables $\bm{x}_t=(1,x_{1t},...,x_{5t})$
are generated by following the structure from \cite{george1997approaches}. First we generate six independent standard normal random variables, $z_0,...,z_5$. We generate five correlated covariate variables by $x_{1t} = 0.4z_0+z_1$, $x_{2t}=x_{1t}+1.8z_2$, $x_{3t}=0.4z_0+ z_3$,
$x_{4t} = x_{3t} + 1.4z_4$, and $x_{5t} = x_{3t} + 1.4 z_5$, resulting in the covariate vector $\bm{x}_t$ that are independent across time $t$. 
The dependent variable $y_t$ is generated by $y_t = \bm{x}'_t\bm{\beta}_t + \epsilon_t$,
where the error term $\epsilon_t$ is independently drawn from a standard normal distribution. The coefficients, $\bm{\beta}$, 
are chosen such that some of the parameter values abruptly changes between different regimes of constant values or zero. In particular, $\beta_0$ is set to zero, rendering a zero intercept. $\beta_2$ and $\beta_5$ are both zero across all time points, rendering zero impact from the covariates $x_{2t}$ and $x_{5t}$. $\beta_1$ has 3 regimes, with the parameter values being constant but non-zero in each regime. $\beta_3$ has 3 regimes, with the parameter taking non-zero constant values in the first two regimes before switching to a zero value in its last regime. $\beta_4$ also exhibit 3 regimes, but with the second of three regimes being the zero-state, while the other two being non-zero constants, implying that the impact $x_{4t}$ is switched on and off.
The dynamic of the true parameter values are shown in the green dashed line in the top and middle panels of Figure~\ref{fig:simu_msdsp_beta}

Figure~\ref{fig:simu_msdsp_beta} depict the comparison of posterior inference of the TVP between the DSP 
and the MSDSP models. The top panel shows the posterior means (solid yellow line) 
and point-wise 95\% density intervals (gray shade) inferred from the DSP model.
The middle panel depicts the posterior means (solid yellow line) 
and point-wise 95\% density intervals (gray shade) inferred from the MSDSP model.
The bottom panel plots the true Markov switching state $\bm{s}_t$ (green dashed line)
and the posterior probability of being in the non-zero DSP-state (yellow solid line).
In inference of both the DSP and the MSDSP models, we use identical uninformative prior specification for the parameters related to the DSP component for comparability.

From the bottom panel in Figure~\ref{fig:simu_msdsp_beta},
the true state and the inferred state probability almost coincide, which indicates an almost perfect ability of the MSDSP model to switch from the zero state to the DSP state.
Comparing the top and middle panels,
we observe that when a coefficient's true value is in the zero-state,
the MSDSP will quickly adapts to that state.
This is particularly evident in the inference of $\beta_0, \beta_2$ and $\beta_5$, where 
the posterior uncertainty is much smaller for the MSDSP. Since the DSP is unable to shrink to to absolute zero-state, the the posterior inference over these true zero-states still exhibit a larger degree of uncertainty, even when the posterior means are hovering around zero.
Similarly, the zero-state episodes for $\beta_3$ and $\beta_4$ are well captured with much tighter posterior intervals when the MSDSP is used.
For non-zero coefficient values, the MSDSP and DSP performs similarly, with both models able to capture the shifts of the coefficients to a non-zero constant value effectively.

\subsection{Case 2: Gradual Shifts in TVP}
\begin{figure}
    \centering
\includegraphics[width=1\textwidth]{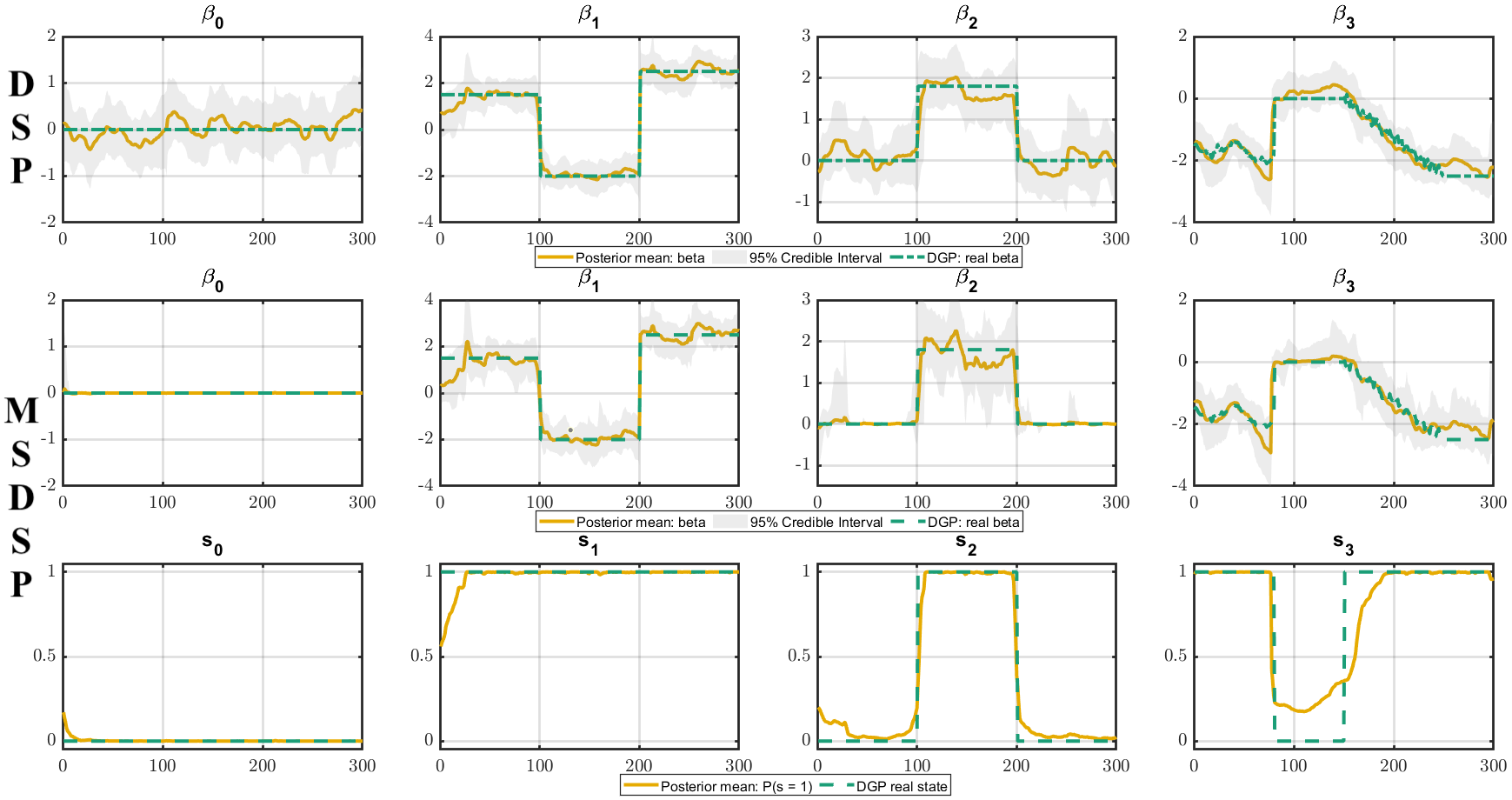}
\caption{Posterior estimates from DSP and MSDSP models on DGP 2. 
The first two rows show the posterior mean and 95\% credible intervals of 
$\bm{\beta}_t$ from DSP and MSDSP, respectively. 
The green dashed lines are the true values.
The bottom row shows the posterior probabilities of being in the DSP-state 
in the MSDSP model. The green dashed lines are the indicators of the true DSP-state.}
     \label{fig:simu_msdsp_beta_p4}
\end{figure}

For this case, we simulate three explanatory variables, $\bm{x}_t=(1,x_{1t},...,x_{3t})$, from independent standard normal distributions with $T=300$.
As in Case 2, $\bm{x}_t$'s are independent across time $t$.
The coefficients, $\bm{\beta}$, are given in Figure~\ref{fig:sim_true}
and described in the Introduction. They are also shown as the green dashed lines in the top and middle panels 
in Figure~\ref{fig:simu_msdsp_beta_p4}. Again, the dependent variable $y_t$ is generated by
$y_t = \bm{x}'_t\bm{\beta}_t + \epsilon_t$,
where the error term $\epsilon_t$ is independently drawn from a standard normal distribution.

This collection of coefficients encompasses all possible scenarios that the dynamic TVP, and we compare the DSP and MSDSP models in this setting. We highlight that the difference between Case 1 and Case 2 is that Case 2 exemplifies the gradual change in parameter in the scheme for $\beta_3$.
As in Figure~\ref{fig:simu_msdsp_beta} for Case 1, Figure~\ref{fig:simu_msdsp_beta_p4} depicts the posterior mean and point-wise 95\% posterior interval of $\bm{\beta}_t$, with those produced by the DSP model depicted in the top panel, and those from the MSDSP depicted in the middle panel. The bottom panel depicts the posterior summary for the Markov switching variable $\bm{s}_t$. Consistent with what we observed in Case 1, when the coefficient enters the zero-state, the MSDSP model is able to capture this scenario with a much higher degree of certainty.
When the parameters are non-zero, the MSDSP and DSP models perform similarly, including when the shift in the parameter is gradual, as in the case of $\beta_3$.

\section{Addressing the Meese-Rogoff Puzzle with MSDSP}\label{sec:app}

We employ the flexible MSDSP model to offer a novel econometric perspective on the Meese-Rogoff puzzle. In particular, we seek to establish if the model's ability to achieve sparsity and incorporate dynamic parameters that can potentially shrink toward a constant makes economic models competitive with random walk predictions. To set notation, let $e_t$ denote the logarithm of spot change rate between a country to the US dollar at time $t$, 
and we construct the response variable as $y_t = e_t - e_{t-1}$ 
as the growth rate of exchange rate at time $t$.
We use the direct quotation method, hence a positive $y_t$ means
that the domestic currency depreciates against the US dollar.

The Meese–Rogoff Puzzle \citep{meese1983empirical, meese1983out, meese1988was} 
claimed that the random walk model is a competitive model 
that is difficult to beat for exchange rate prediction.
Therefore, we view the random walk as the benchmark model and augment it with 
a SV component to reflect the contemporary development 
in the empirical macroeconomic modeling literature.
Specifically,
 \begin{align}  
    y_t &= \mu + \epsilon_t, \qquad \epsilon_t \sim N(0, \exp(g_t)),  \label{eq:rw_y}\\
     g_t &= \mu_g + \phi_g (g_{t-1} - \mu_g) + e_t, \qquad e_t \sim N(0, \sigma_g^2). \label{eq:rw_g}
\end{align}  
If we restrict the $\bm{x}_t=1$ in the measurement equation, \eqref{eq:msdsp_y}, and assume a constant intercept, the MSDSP model reduces to the random walk model. We use both the random walk model with non-zero drift (\textbf{RW-drift-SV}) and the driftless random walk where $\mu=0$ (\textbf{RW-SV}). As pointed out by \cite{rossi2013exchange}, the RW-SV is the toughest benchmark to beat. Since both RW-drift-SV and RW-SV models contains no external covariates, we apply the indirect method to recursively construct the $h$-period ahead forecast from these models.

\subsection{Competing Economic Models with MSDSP} \label{subsec:eco_models}
We respect the existing literature, as summarized by \cite{rossi2013exchange},
to include four different models driven by economic theory, along with three further predictive models based on
commodity prices (oil, gold, and copper)
inspired by \cite{ferraro2015can}. All seven competing economic models can be written in the form of a linear model in \eqref{eq:y_th}, but with varying set of covariates. We summarize the specification of the covariate dictated by each of the economic model in Table \ref{tab:econ_models}.

\begin{table}
    \centering
    \begin{tabular}{lll}
        Model & Acronym  & $\bm{x}_t$ \\ \hline \hline
         Interest Rate Parity& IRP  &  \( (1, (r_{t} - r_{t}^*)  )\) \\
         & & \(r_{t}- r_{t}^*=\) difference in log interest rates \\ \hline
         Taylor Rule & TR & \((1, (r_{t-1} - r_{t-1}^*) , \left[ 1.5 (\pi_{t} - \pi_{t}^*) + 0.5 (x_{t} - x_{t}^*) \right] )\)  \\
         & & \(\pi_{t} - \pi_{t}^*=\) difference in inflation rates \\
         & & \(x_{t} - x_{t}^*=\) difference in output gaps \\ \hline
         Purchasing Power Parity& PPP  & \((1,(p_{t} - p_{t}^* - e_{t}))\) \\
         & & \(p_{t} - p_{t}^*=\) difference in log price levels \\
         & & \(e_t=\) exchange rate at time $t$\\ \hline
         Monetary Model & MON  & \( (1,\left[ (m_{t} - m_{t}^*) - (g_{t} - g_{t}^*) - e_{t} \right] )\)\\
         & & \( m_{t} - m_{t}^*=\) difference in log money supply\\
         & & \(g_{t} - g_{t}^*=\) difference in log real output \\ \hline
         Commodity Models&  Oil& \((1,(p^{oil}_{t} - p^{oil}_{t-1}))\) \\
         & Gold & \((1,(p^{gold}_{t} - p^{gold}_{t-1}))\)\\
         & Copper & \((1,(p^{copper}_{t} - p^{copper}_{t-1}))\)\\ 
         & & \(p_t^{x}\) denotes the log price of commodity `$x$' at time $t$\\\hline \hline
         \end{tabular}
    \caption{Summary of the competing economic models. The model specification follows \eqref{eq:msdsp_y}-\eqref{eq:msdsp_s}, with the set of \(\bm{x}_t\) changing according to the economic model framework. All quantities denoted with `*' corresponds to those of the U.S.}
    \label{tab:econ_models}
\end{table}

The adoption of MSDSP (and its nested DSP) in the time-varying coefficient in these models allow for the coefficient to move over time, as well as allow for stochastic volatility in the predictive distribution of exchange rate changes. Both features in the econometric modelling of these economics theories have been tested in
the contemporaneous macroeconomic literature.
We employ both the MSDSP and the DSP specifications on all seven competing economic models listed in Table \ref{tab:econ_models}. Our analysis provide a new perspective to the Meese-Rogoff puzzle, with the analysis allowing for additional flexibility in the economic models to allow them to adapt to changing economic environments. 

The two random walk models, RW-SV and RW-drift-SV, along with the constant coefficient versions of the seven competing economic models, serve as benchmarking models in our analysis. We refer to the constant coefficient economic models simply as ``linear'' models. For the linear economic models, 
we consider expanding window as well as moving window of 100 months and 50 months
to capture potential structural instability brutally.
All models, including the RW-SV, RW-drift-SV and the linear benchmark models, are inferred in the Bayesian framework.
Hence, we take account the parameter uncertainties and
are able to provide a full profile of the predictive distribution,
including densities and point forecasts.


We evaluate our predictive distributions over the forecast horizon $h=1,...,12$. Since we employ monthly data, these horizons corresponds to from 1 month to 1 year.
All predictions are out-of-sample by using expanding as delineated in 
Section~\ref{sub:oos}, unless stated otherwise. For completeness, we also report the contemporaneous forecasts
conducted in \cite{ferraro2015can}, which is equivalent to 
predicting $y_{t+h}$ by assuming $x_{t+h}$ exists, in the appendix.
\footnote{
This approach is a pseudo prediction since it ignores simultaneity.
It is not a fair comparison to the random walk model,
but can provide information for comparison between economic models
by stripping off predictor's uncertainty.}

\subsection{Forecasting Metrics}
\label{subsec:metrics}
We consider \textbf{five} metrics to evaluate the predictive distributions of the competing models:
log predictive likelihood (Bayes factor),
continuous ranked probability score (CRPS),
root mean squared forecast error (RMSFE),
tail coverage rate,
and model confidence set (MCS).
All metrics are out-of-sample and calculated based on the Monte-Carlo estimate of \eqref{eq:postpred}, as described in Section~\ref{subsec:oos}. In what follows, we define $T_0$ as the initial sample period, and $T$ denotes the overall sample size, with the number of $h$-period ahead periods evaluated being $T-h-T_0+1$.

First, the predictive distribution is evaluated by the log predictive density ratio (LPDR), which evaluates the predictive performance of model $\mathcal{M}$ relative to the benchmark model $\mathcal{M}_B$:
\begin{align}
\textbf{LPDR}(\mathcal{M}, \mathcal{M}_B) = \log(p(Y_{T_0+h:T}\mid I_{T_0}, \mathcal{M}))-\log(p(Y_{T_0+h:T}\mid I_{T_0}, \mathcal{M}_B)).
\end{align}
Here, \[\log(p(Y_{T_0+h:T}\mid I_{T_0}, \mathcal{M}))=\sum_{t=T_0}^{T-h}\log(p(y_{t+h}\mid I_t, \mathcal{M}))\] is the cumulative log score, with  \(\log(p(y_{t+h}\mid I_t, \mathcal{M}))\) denoting the log predictive density of the $h-$period ahead prediction constructed based on information set up to time $t$. The LPDR is promoted by \cite{Geweke2010-kb}, and can be viewed as the log predictive Bayes factor of model \(\mathcal{M}\) relative to the benchmark \(\mathcal{M}_B\). A value that is larger than $3$ means that model $\mathcal{M}$
is strongly preferred to the benchmark model $\mathcal{M}_B$ (see \citep{Kass1995-ke}). With the RW-SV being identified by \cite{rossi2013exchange} as being the toughest model to beat, the RW-SV model serves as our $\mathcal{M}_B$ for all subsequent analyses.

The \textbf{continuous ranked probability score} (CRPS)
assesses the full predictive distribution prioritizing the sharpness of the distribution and is thus less sensitive to tail behaviors
\citep{gneiting2007strictly}. It is defined, for model $\mathcal{M}$, as
\begin{align}
\textbf{CRPS}_{t+h}\left(p(y_{t+h}\mid I_t, \mathcal{M}), y_{t+h}\right)
=\int_{-\infty}^{\infty} \left(\bm{1}\left(y_{t+h} \leq r\right)-F(r\mid I_t, \mathcal{M})\right)^2\mathrm{d}r, \nonumber
\end{align}
where $F(r\mid I_t, \mathcal{M})=\int_{-\infty}^r p(\tilde{y}_{t+h}\mid I_t, \mathcal{M}) d\tilde{y}_{t+h}$
is the CDF for $y_{t+h}$ and
$\bm{1}(.)$ is the indicator function which returns one when the condition in the parenthesis is true. 
We report the average CRPS 
\begin{align}
\textbf{CRPS}_{T_{0+h}:T}(\mathcal{M}) = \frac{1}{T- T_0 -h +1} \sum_{t=T_0}^{T-h} \textbf{CRPS}\left(p(y_{t+h}\mid I_t, \mathcal{M}), y_{t+h}\right).
\end{align}
for model comparison.
A smaller value means better performance.

For point forecast evaluation, 
the \textbf{root mean squared forecast error} (RMSFE) is one standard approach, 
and computed as follows:
\begin{align}
\text{RMSFE}_{T_{0+h}:T} = \sqrt{\frac{1}{T- T_0-h+1 } \sum_{t=T_0}^{T-h} \left(y_{t+h}-E(y_{t+h}
\mid I_t, \mathcal{M})\right)^2},
\end{align}
where the conditional mean $E(y_{t+h}\mid I_t, \mathcal{M})$
is calculated by the sample average of the corresponding set of $\{y_{t+h}^{(g)}\}_{g=1}^G$.

To evaluate the tail behavior, we report the \textbf{coverage rate} (CR)
at different tail quantiles.
It is the ratio of the number of observations falling into a 
specified range to the total number of observations.
For example, to assess the predictive 5\% quantile,
the coverage rate will be the number of observations that are below the predictive 5\% quantile value
to the total number of observations.
Obviously, a model with a ratio closer to 5\% would be favored in this example.
In general, we define, for the lower-tailed $\alpha$\% quantile prediction,
\begin{align}
\textbf{CR}(\alpha,\mathcal{M})=\frac{1}{T-T_0-h+1} \sum_{t=T_0}^{T-h} 
\bm{1}\left(y_{t+h} \leq q^{\alpha,\mathcal{M}}_{t+h}\right),
\end{align}
where $q^{\alpha,\mathcal{M}}_{t+h}$ is the predictive quantile from the predictive distribution, such that $Pr(y_{t+h}\leq q^{\alpha,\mathcal{M}}_{t+h}|I_t,\mathcal{M})=\alpha$.
Analogously, it is straightforward to calculate the upper-tailed quantile coverage rate, with the ``$\leq$'' sign replaced with the ``$\geq$'' sign in all calculations above.

Lastly, we construct the \textbf{Model Confidence Set} (MCS, \citep{Hansen2011-hy}, and 
executed in \citep{MSC_R})
to select a list of `superior' models. 
The MCS controls for multiple comparisons to avoid inflated Type I error and is robust to model mis-specification, 
especially when evaluating out-of-sample predictive performance.
We implement the MCS using three different out-of-sample loss functions for MCS: Squared Forecast Errors (SFE),
negative Log Score, and Continuous Ranked Probability Score (CRPS).
The MCS produces a set of ``competitive'' models within the given confidence set along with their respective rankings, with uncompetitive models discarded.
We choose a 95\% model confidence set in our application.

\subsection{Data Description} \label{subsec:dt}

With the United Kingdom and the USA being key international financial market players, we present our application using the United Kingdom's British Pound (GBP/USD) exchange rate and discuss the results comprehensively. We also analyze the cases ofthe Canadian dollar (CAD/USD) and Japanese Yen (JPY/USD) exchange rates, but their results are provided in the appendix due to the volume of results from the variety of models investigated.

Following \cite{rossi2013exchange} and \cite{Aristidou2022-ox, Aristidou2022-pg}, 
we collected data from the following primary sources: 
FRED (Federal Reserve Economic Data), 
the IMF (International Monetary Fund), 
the OECD (Organisation for Economic Cooperation and Development), 
and the Bank of England. We employ the end-of-month nominal exchange rate for GBP/USD. 
As noted by \cite{rossi2013exchange}, end-of-month data generally exhibits weaker predictive power, providing a lower bound for forecasting accuracy. 
The monthly total industrial production is used as a proxy for real GDP,
while the real potential GDP is derived by using the one-sided HP-filter. 
The other times series include the Producer Price Index (PPI), 
monetary aggregates (M1), 
and commodity prices for crude oil, gold, and copper. 
All variables are standardized to have a mean of 0 and a variance of 1.
Detailed data sources and series codes are provided in section A of the supplementary material.

Due to the availability of commodity prices and interest rates, 
the longest common sample period spans from January 1990 to June 2017, 
totaling T=330 months. The out-of-sample period is from Nov 2000 to June 2017, 
comprising 200 observations, leaving the initial sample period $T_0=130$ observations.

\subsection{Analysis of Predictive Performance}\label{sub:oos}

We first summarize that the economic models with the MSDSP flexibility perform the best in every out-of-sample evaluation metric.
It consistently demonstrates superior forecasting performance compared to the random walks, producing higher LPDR,  lower CRPS,
lower RMSFE, more accurate coverage rate, as well as always attaining high ranks in the MCS assessment. 
At the same time, our results are also consistent with the literature, in that the random walk models are superior to the linear economic models (expanding or rolling window)
in general. While there are a few exceptions to this general conclusion, but they are not systemic do not alter our overall conclusions that the added flexibility from the MSDSP prior indeed lead to better predictive performance of the economic models. Our observations also extend to the cases of Canadian dollar and Japanese Yen, with their detailed results provided in sections D and E of the supplementary material, respectively. In what follows, we discuss the GBP/USD results in more details.  

\subsubsection{Density Forecasts}
Table~\ref{tab:UKlpdr13612} reports the LPDR of the competing models, evaluated relative to the RW-SV benchmark, at selected horizons of
$1,3, 6$ and $12$ months. In each row, the bolded statistic indicated the best predictive score, with positive numbers indicating that particular model outperforms the benchmark. It is evident that the MSDSP version of each economic model is always the best performing, with the MSDSP column always bolded. The results for other forecast horizons are provided in section C of the supplementary material and are consistent with those reported in Table~\ref{tab:UKlpdr13612}.

Importantly, the first three columns under ``Linear'' always produce negative LPDR, indicating that the basic economic models are always worse than the random walk,
an observation that is consistent with the Meese–Rogoff Puzzle.
Once the MSDSP is implemented, however, 
the results are overthrown.
This is the decisive evidence that
MSDSP can discover the nonlinearity and dynamic instability to
improve exchange rate prediction.
In addition, the predictive LPDR from the DSP column are all negative as well, highlighting the value added from the Markov switching process that we proposed, suggesting that the DSP itself is not parsimonious enough to produce a competitive density prediction relative to the RW-SV.

The above interpretation of results in Table~\ref{tab:UKlpdr13612}
is convincing, and more than sufficient to provide evidence for the predictive dominance of our MSDSP framework. Convincingly, the MSDSP outperforms the RW-SV at all horizon and with all economic models. Based on the magnitude of the LPDR itself, we can also infer the degree of the superiority of the predictive performance. For example, when $h=1$, the best performing model
is the MSDSP-PPP, which is the time varying parameter PPP model
with an MSDSP prior. The LPDR is larger than $3$ and hence it is strongly supported by data
against the random walk model RW-SV.
In fact, the MSDSP-PPP model strongly dominates the random walk at
all horizons, providing a strong signal that
MSDSP captures important data dynamic features. These observations are also confirmed by the analysis of CRPS metric, reported in Table~\ref{tab:ukcrps13612}, with relative differences in this metric also confirming the superiority of the MSDSP models relative to the RW-SV, as well as its nested alternative the DSP specification.

From the angle of tail prediction, we report the 1-period ahead coverage rate
for top and bottom $2.5\%$ and $10\%$, respectively, in Table~\ref{tab:ukcp02510}\footnote{Due to page limitation, other horizon's results are available upon request.}. 
The MSDSP does not perform ideally in bottom $2.5\%$ and top $10\%$, 
but it is much better than the RW-SV in all other quantile regions. The MSDSP models also generally produces more accurate coverages than their linear counterparts. We observe that the RW-SV tends to provide tail predictions that produce under-coverage.
For example, in the top panel (Top $2.5\%$)
of Table~\ref{tab:ukcp02510}, the coverage rate 
from MSDSP is very consistent to $2.5\%$, but the random walk
only covers $1.5\%$.
In the third panel (Top $10\%$), 
the coverage rate from MSDSP is about $8\%$,
but the random walk benchmark is only $4.5\%$.
In summary, the MSDSP still performs better than the RW-SV.

\subsubsection{Point Forecasts}
The root mean squared forecast error (RMSFE)
is the most common evaluation prediction metric for point forecasts, 
especially in the frequentist approach.
The RW-SV model usually has its home advantage in this metric \citep{rossi2013exchange}.
We are excited to reveal that the MSDSP performs extremely well with point forecasts according to the RMSFE. Table~\ref{tab:ukrmsfe13612} shows the ratio of 
RMSFE of the competing models relative to that of the RW-SV benchmark model.
A value less than $1$ means that the corresponding model produces smaller RMSFE statistic than the RW-SV benchmerk.

With only one exception, the MSDSP model produces smaller RMSFE statistics compared to the RW-SV model.
The only exception is the linear economic model using copper prices, based on a 50-period rolling window scheme. Even in this case, the MSDSP method is still
better than the RW-SV, and is only slightly worse than the linear counterpart. As in the case of density forecast, the MSDSP dominates the RW-SV benchmark, but the RW-SV is better than linear or DSP economic models in most cases. We also observe that the linear models based on commodity prices are also capable of performing better than the RW-SV model in point forecasting, although this cannot be generalized to density predictions.

We carried out pairwise Diebold-Mariano (DM) tests for all the 
models against the RW-SV. Unfortunately, we did not achieve many significant result despite the MSDSP column in Table~\ref{tab:ukrmsfe13612} indicating the direction for better performance.
We also performed the Wilcoxon signed-rank test, which is a joint test based on the DM statistics from 
for all pairs of MSDSP models and the RW-SV benchmark ($7\times 12=84$). The joint test reports the p-value of $0.0000$, indicating that there is indeed a significant difference between the RMSFE of the MSDSP models compared to the RW-SV.
We also performed the same Wilcoxon signed-rank test
for each linear economic model with its MSDSP counterparts ($12$ statistics/horizons each).
For all variants of the model, the test reports p-values less than $0.001$, with exception of the model that involve gold prices returning 
a p-value of $0.017$. This reflects a strong evidence that the superior performance of the MSDSP is supported by the data, even for
point forecast.

\subsubsection{Model Confidence Set and Rankings}
Table~\ref{tab:ukmcs_h12} reports the $95\%$ Model Confidence Sets 
for $h=1$ and $h=2$ months ahead forecasts based on three evaluation criteria: 
Squared Forecast Error (SFE), negative Log Score, and the Continuous Ranked Probability Score (CRPS).
Each column report integer rank of each of the model included in the $95\%$ model confidence set, 
for the corresponding selection criterion and forecasting horizon. 
Symbol ``-'' indicates that the model is eliminated form the confidence set, which implies that the models are not ranked.

The MSDSP is always the number $1$ choice in the confidence sets.
With only two exceptions, the MSDSP is always among top-10 choices.
Considering there are 7 MSDSP economic models, this is a striking result.
Amazingly, at horizon $h=1$, all three criteria
admit that MSDSP method dominate any other models, 
since the seven MSDSP economic models ranks the top 7 in each of the model confidence sets.
Although the RW-SV and RW-drift-SV models are not rejected and are retained in the confidence sets,
their highest ranking is 8. 

Once again, we observe that the economic models, along with the flexibility of the MSDSP, beats the RW-SV benchmark.
Our MCS results also confirm prior discussions on the DSP model performance being inferior, and are often eliminated from the confidence set, highlighting the added value of  Markov switching mechanism 
in exchange rate prediction. We also confirm the superiority of the RW-SV and RW-drift-SV compared to the linear economics models, with the benchmark models' rankings consistently higher than those of the linear models.

\section{Model Assembly with MSDSP}\label{sec:assembly}

Since there has been weak consensus on the dominant model choice in exchange rate forecasting,
a natural path is to combine the predictions via model assembly.
Researchers applied the Bayesian Model Averaging (BMA) 
\citep{wright2008bayesian, della2009economic}
or dynamic Bayesian Model Averaging (DBMA) \citep{vcasta2024forecasting}
in exchange rate forecasting, combining many weak models to form a strong prediction.
We do not intend to exhaustively explore the model combination method 
due the space limitation and our focus on MSDSP.
We consider each prediction from a small model can be viewed as an expert's opinion, and propose a new model assembly method that naturally 
adapts to the MSDSP setting
based on dynamic Bayesian predictive synthesis (DBPS) from \cite{mcalinn2019dynamic}. The adaptation of the MSDSP allows for the predictive synthesis to dynamically adapt, with shrinkage of the combination parameters to a constant or a switch to no contribution in a single framework. We investigate if the use of MSDSP in this setting also leads to improved predictive performance in exchange rates.


\subsection{Dynamic Bayesian Predictive Synthesis}
\label{subsec:dbps}
We build our method on the dynamic Bayesian predictive synthesis from \cite{mcalinn2019dynamic}.
There is a burgeoning literature as \cite{johnson2017bayesian,McAlinn2020-ke,Aastveit2023-wv,Tallman2024-mu}.
The DBPS of \cite{mcalinn2019dynamic} can be expressed as follows
 \begin{align}
 y_t&=\omega_{t, 0}+\sum_{j=1}^L \omega_{t, j} \hat{y}_{t, j}+e_t, \quad &e_t& \sim N\left(0, v_t\right), \label{eq:dbps:y}\\
 \omega_t&=\omega_{t-1}+w_t, \quad &w_t& \sim N\left(0, v_t W_t\right),
\end{align}
where $v_t$ is defined via a standard beta-gamma random walk volatility model \citep{Prado2010-ec} 
and $W_t$ follows a standard single discount factor specification \citep{West1999-ic}. 
Each $\hat{y}_{t, j}$ is the prediction from model (or expert) $j$.
The dynamic BPS considers distribution input, 
hence $\hat{y}_{t, j}$ is treated as random.
The vector $\omega_t=(\omega_{t, 0},\omega_{t, 1},...,\omega_{t, L})'$
is the time-varying intercept and weights, with $L$ here denoting the number of models or experts.
The intercept $\omega_{t, 0}$ captures the systemic time-varying bias.
There is no restrictions to the model weights $\omega_t$.
The absence of any restrictions provides a great deal of flexibility.
For example, it allows a ``bad'' model
to provide ``good'' results, as a negative weight is also permitted.
It can also take advantage of a systemically upward biased model 
by giving it a discount,
resulting the sum of model weights being less than unity.

The conditional posterior distribution of 
each model's prediction $\hat{y}_t=(\hat{y}_{t, 1}, ..,, \hat{y}_{t, L})'$ 
is updated as
is given by:
 \begin{align*}
p\left(\hat{y}_t \mid Y_{1: t}, \omega_t, M_{1: L}\right) \propto \alpha\left(y_t \mid \omega_t, \hat{y}_t\right) \prod_{j=1}^L p\left(\hat{y}_{t, j} \mid y_{1: t-1}, M_j\right)
\end{align*}
where $ \alpha\left(y_t \mid \omega_t, \hat{y}_t\right)$ is 
the density implied by \eqref{eq:dbps:y},
which is named as \emph{synthesis function}. 
The predictors $\hat{y}_{t, j}$ are sampled from 
this conditional kernel. 
This sampling approach can be considered 
as tilting the originally standalone predictive distribution of model $j$ 
by the observations via the synthesis function and
associated interactions with other models.

Conditional on a draw of $\hat{y}_t$,
standard MCMC can be applied.
To ensure reliability and consistency, 
both inference and prediction procedures 
are implemented by using the code provided on McAlinn's website.
The hyper-parameters of the DBPS such as the discount factor
are set the same as \cite{mcalinn2019dynamic}.
We have also performed robustness checks on these hyper-parameters and results are available upon request.

\subsection{Application of MSDSP and DSP}
The idea of DBPS is very appealing to practitioners.
But we do not want to increase additional computational cost on 
simulating $\hat{y}_t$.
Hence, instead of inputting predictive  distributions,
we only input one prediction value of $\hat{y}_t$ such as the predictive mean
and treat it as the data.
This would free the cost of computation by
make the assembly a two-stage task.
In stage 1, we compute the prediction from each model.
In stage 2, assemble these models by the synthesis function.

Our approach is closer to the pooling method 
such as static pooling \citep{Hall2007-ye,Geweke2011-ap} 
and dynamic pooling \citep{Waggoner2012-vb,Del-Negro2016-gz}. 
Some other variants include two-state Markov process pooling 
by \cite{Waggoner2012-vb} and 
the infinite-state Markov process pooling by \cite{Jin2022-gj}.
There are two key differences.
First, we use the MSDSP as the weight dynamics,
so that sparsity is explicit.
Second, we use the DBPS idea to incorporate 
the bias term in the synthesis function.
Our method is also close to
ensemble learning via stacking in machine learning 
\citep{WOLPERT1992-241, Breiman1996, Dietterich2000, 
Proscura2022, MUSLIM2023-200204}.

The specifics are expressed as follows.
\begin{align}
 y_t&=\omega_{t, 0}+\sum_{j=1}^L \omega_{t, j} \hat{y}_{t, j}+e_t, \quad e_t \sim N\left(0, e^{v_t}\right), \label{eq:assembly:msdsp:y}\\
 v_t &=\mu_v + \phi_v(v_t-\mu_v) + e^v_t, \quad e^v_t\sim N(0, \sigma^2_v)
 \label{eq:assembly:msdsp:v}
\end{align}
where $\omega_{t, 0}$ is the time-varying intercept 
and $\omega_{t, j}$, for $j=1,...,L$, is the time-varying 
weight for $j$-th model's prediction as in DBPS.\footnote{
We double use the notation $\omega$, 
which also exists in \eqref{eq:msdsp_beta_tilde}, to represent the model weights.
Since Equation~\eqref{eq:msdsp_beta_tilde} is not explicit in 
this section, we hope readers can understand that these two $\omega$'s
are different.  
}
The difference in \eqref{eq:assembly:msdsp:y}
is that we treat $\hat{y}_{t,j}$ as data
in MSDSP assembly.
In this paper, we use out-of-sample
predictive mean as $\hat{y}_{t,j}$ while other options such as the median is also feasible.
One can also input multiple predictive values from one model
to exploit various information content in our framework,
because the computation is parallelizable and the sparsity is explicit.
The volatility of $y_t$
is set as the standard SV in \eqref{eq:assembly:msdsp:v}.
The dynamic of each $\omega_{t,i}$, for $i=0,1,...,L$,
is set as an independent MSDSP process in \eqref{eq:msdsp_beta}-\eqref{eq:msdsp_s}.

For simplicity, we set $L=7$ base models
as the linear economic models in Section~\ref{subsec:eco_models}. 
For prediction at horizon $h$, 
simply replace $\hat{y}_{t, j}$ by $\hat{y}_{t+h, j}$
and $y_t$ by $y_{t+h}$, respectively. As long as the information set used to produce these predictions is up to time $t$, these
are out-of-sample prediction. In addition to the MSDSP assembly, we also apply its nested alternative, the DSP, in model assembly, along with the standard BPS framework described in Section~\ref{subsec:dbps}, for comparison.
Note that no MSDSP or DSP methods are used in the construction of our base models. The ideas applying ``MSDSP-in-MSDSP'', that is ensembling MSDSP models with MSDSP weights are left for future research.

\subsection{Predictive Scheme with Model Assembly}

Because of the the two-stage procedure for model assembly,
we cut the data at two time points: $T_0$ and $T_1$ with $T_0<T_1<T$.
The initial training sample uses the information up to time $T_0$,
which is the same as in Section~\ref{sec:app}.
The initial training sample serves to construct the predictions from the base models in this section.
The period between $T_0$ and $T_1$
is used to train the synthesis function, and thus the structure of the model weights $\omega_t$
It cannot be inferred without input from the base models,
so the two-stage method wins us some computational time,
at the cost of a smaller testing sample.
We choose $T_0$ and $T_1$ such that $T-T_1=100$ and $T_1-T_0=100$, and keep $T_0$ the same as in Section~\ref{sec:app}.
This idea of breaking the sample to three subset is analogous to the training, validation and testing sampling 
scheme in the machine learning literature, 
but we keep updating the data after each prediction 
and hence can exploit the information as much as possible.

\subsection{Model Assembly Results}

Table~\ref{tab:ukbps} reports 
the log predictive likelihood (LPL),
Continuous Ranked Probability Score (CRPS), and Root Mean Squared Forecast Error (RMSFE)
for horizons $h=1,...,12$ months.
These values are based on the last $100$ samples, adjusted by the forecasting horizons.
The methods include the MSDSP and DSP assembly, along with the standard DBPS of \cite{mcalinn2019dynamic}. The RW-SV predictive metrics over the corresponding sample period are also reported for comparison.
Note that all base models are the linear economic models with static parameters for both versions of model assembly and DBPS.

Amongst the model assembly method, the MSDSP performs better than the DSP assembly and the standard DPBS in all evaluation metrics at all horizon. 
Thus, we show here that the MSDSP assembly can easily dominate the state-of-the-art dynamic Bayesian predictive synthesis.
Without the need to deploy heavy computation for
the distributional input, the MSDSP assembly is scalable.
Its Markov switching structure explicate sparsity to improve prediction,
which is evidenced by the dominance of MSDSP assembly over the DSP assembly.

The MSDSP assembly is slightly worse than the random walk at short horizons
but tends to improve at longer horizons.
For example, the point forecast (RMSFE)
from MSDSP is better than the random walk when $h>9$,
which is empirically plausible (see also \citealp{mark1995exchange,
rapach2002testing, molodtsova2009out}). A better model pool might improve the MSDSP assembly,
but it is not our focus in this paper.

\section{Discussion and Conclusion}
\label{sec:concl}
We propose the novel Markov Switching Dynamic Shrinkage Process method
to address dynamic instability, model parsimony and sparsity, and use the proposed framework to address the renowned Meese–Rogoff puzzle,
which states that the random walk is highly competitive in exchange rate forecasting. We consider adopting the MSDSP in two settings: directly incorporating the flexible structure into existing economic models; and applying the flexibility in model assembly of standard linear economic models. 

In our application, we show that adopting the MSDSP directly to enhance the economic model structure leads to significant improvements in the predictive performance of the economic models relative to the benchmark random walk. The decisive evidence illustrate that the random walk model, even when adjusted to include stochastic volatility, is systematically inferior to the economic models improved with MSDSP priors for both point and density forecasts. This evidence, in itself, provide a new perspective on the Meese-Rogoff puzzle, highlighting the fact that the models resulting from economic theories possess predictive power when flexibility and sparsity is taken into account. Existing results literature have evaluated these models in its most stringent form via the constant coefficient linear model, under the assumption that the respective economic relationship is static, which, in the current economic environment, is unrealistic. 

Furthermore, we provide a natural framework for model assembly 
by using the MSDSP toolkit.
With a very primitive linear constant coefficient model pool, the MSDSP assembly consistently beat the state-of-the-art dynamic Bayesian predictive synthesis,
while being scalable computationally. The MSDSP model assembly produces similar predictive performance as the benchmark random walk. Compared to the individual model improvement with MSDSP, there is still further improvements that can be investigated with future research, including the inclusion of more sophisticated base model and the incorporation of predictive distribution (rather than point) in the construction of predictive synthesis.

\newpage 
\section{Tables}
\begin{table}[!htbp]
\centering
\caption{UK: Comparative Forecast Performance (1,3,6,12-period-ahead forecasting) Using Log Predictive Density Ratio (LPDR) with reference: Random Walk with Stochastic Volatility with No Drift. Out-of-sample period: 2000/11 - 2017/6, in total 200 periods.}
\label{tab:UKlpdr13612}

\begin{tabular}{lrrrrr} 
\toprule 
  & \multicolumn{3}{c}{Linear} & MSDSP & DSP \\
 \cmidrule(lr){2-4}\cmidrule(lr){5-5}\cmidrule(lr){6-6} 
\begin{tabular}[c]{@{}l@{}}Macro\\ Models\end{tabular}  & \begin{tabular}[c]{@{}c@{}}Expanding\\ Window\end{tabular} & \begin{tabular}[c]{@{}c@{}}100-rolling\\ Window\end{tabular} & \begin{tabular}[c]{@{}c@{}}50-rolling\\ Window\end{tabular}  &  &  \\
\midrule
\multicolumn{6}{l}{\textbf{h=1: 1-month ahead.} RW-SV: benchmark. RW-drift-SV: -0.3883} \\
IRP  & -29.6791 & -5.8994 & -5.8353 & \textbf{3.4615} & -9.3826 \\
TR  & -30.5840 & -7.1152 & -6.9684 & \textbf{2.5229} & -13.8571 \\
PPP    & -29.0611 & -6.6741 & -6.6981 & \boxed{\textbf{3.7681}} & -8.6213 \\
MON    & -27.8990 & -4.0341 & -3.9542 & \textbf{2.7240} & -7.6936 \\
Oil  & -26.6876 & -2.0256 & -1.9684 & \textbf{2.2365} & -6.2400 \\
Gold  & -29.2406 & -5.0844 & -5.1214 & \textbf{2.9484} & -10.0253 \\
Copper  & -25.9588 & -0.5934 & -0.6708 & \textbf{2.6266} & -4.5026 \\ 
\midrule
\multicolumn{6}{l}{\textbf{h=3: 3-month ahead. }RW-SV: benchmark. RW-drift-SV: -0.0616} \\
IRP  & -29.1939 & -5.2789 & -5.1703 & \textbf{3.3508} & -8.7248 \\
TR  & -29.7365 & -6.6652 & -6.6531 & \textbf{3.3151} & -14.1063 \\
PPP   & -25.0914 & -3.7528 & -3.6766 & \boxed{\textbf{\textbf{5.2003}}} & -7.4086 \\
MON   & -25.7211 & -3.5646 & -3.6080 & \textbf{3.8906} & -4.9224\\
Oil  & -28.1046 & -3.9143 & -3.8716 & \textbf{3.2649} & -7.2334 \\
Gold  & -28.7496 & -5.3689 & -5.3944 & \textbf{3.0078} & -9.0592 \\
Copper  & -27.4407 & -2.8235 & -2.7777 & \textbf{3.4050} & -6.4234 \\
\midrule
\multicolumn{6}{l}{\textbf{h=6: 6-month ahead. }RW-SV: benchmark. RW-drift-SV: -0.2911} \\
IRP  & -28.2834 & -5.5577 & -5.5701 & \textbf{3.3833} & -8.5100 \\
TR  & -28.2732 & -5.5316 & -5.5212 & \textbf{3.4639} & -12.2628 \\
PPP   & -25.1515 & -4.1046 & -4.0514 & \textbf{3.1787} & -9.3380 \\
MON   & -25.5214 & -4.3304 & -4.3226 & \textbf{2.9245} & -8.7579 \\
Oil  & -25.0861 & -1.8254 & -1.8836 & \textbf{3.0826} & -5.7115 \\
Gold  & -27.5627 & -4.6564 & -4.7751 & \boxed{\textbf{\textbf{3.6933}}} & -8.3479 \\
Copper  & -27.2085 & -3.8037 & -3.9387 & \textbf{3.2561} & -5.9737 \\ 
 \midrule
\multicolumn{6}{l}{\textbf{h=12: 12-month ahead. }RW-SV: benchmark. RW-drift-SV: -0.5118} \\
IRP  & -28.7304 & -4.9605 & -4.8939 & \textbf{3.3245} & -8.2378 \\
TR  & -28.8694 & -5.2189 & -5.2878 & \boxed{\textbf{\textbf{5.9622}}} & -13.3841 \\
PPP   & -24.0625 & -4.1664 & -4.0931 & \textbf{4.0214} & -9.6608 \\
MON   & -24.4873 & -4.0269 & -3.9948 & \textbf{3.7719} & -9.6037 \\
Oil  & -28.1024 & -4.2732 & -4.2203 & \textbf{3.0321} & -9.9155 \\
Gold  & -27.6593 & -3.6843 & -3.7151 & \textbf{3.0996} & -6.4277 \\
Copper  & -28.6770 & -5.4533 & -5.4238 & \textbf{3.0763} & -8.6516 \\ 
\bottomrule
\end{tabular}%
 
\vspace{0.2cm}
\footnotesize
\begin{tabular}{@{}p{\textwidth}@{}}
Note: All LPDR values are computed as the summation over all OOS periods, and recorded as the distance to RW-SV without drift (benchmark values are -272.0914, -272.9213, -272.6078 and -272.7673 for each h). Bold numbers indicate the best-performing model in each row (higher values indicate better predictive performance), and the boxed number indicates the best-performing model in each block. Model abbreviations: IRP = Interest Rate Parity, TR = Taylor Rule, PPP = Purchasing Power Parity, MON = Monetary Model.
\end{tabular}

\end{table} 
\begin{table}[!htbp]
\centering
\caption{UK: Comparative Forecast Performance (1,3,6,12-period-ahead forecasting) Using Relative Continuous Ranked Probability Score (CRPS) - overall score. Out-of-sample period: 2000/11 - 2017/6, in total 200 periods}\label{tab:ukcrps13612}
\begin{tabular}{lrrrrr} 
\toprule 
  & \multicolumn{3}{c}{Linear} & MSDSP & DSP \\
 \cmidrule(lr){2-4}\cmidrule(lr){5-5}\cmidrule(lr){6-6} 
\begin{tabular}[l]{@{}l@{}}Macro\\ Models\end{tabular}  & \begin{tabular}[c]{@{}c@{}}Expanding\\ Window\end{tabular} & \begin{tabular}[c]{@{}c@{}}100-rolling\\ Window\end{tabular} & \begin{tabular}[c]{@{}c@{}}50-rolling\\ Window\end{tabular}  &  &  \\
\midrule
\multicolumn{6}{l}{\textbf{h=1: 1-month ahead.} RW-SV: benchmark. RW-drift-SV: +0.0015} \\
IRP  & 0.0063 & 0.0067 & 0.0071 & \boxed{\textbf{\textbf{-0.0095}}} & 0.0396 \\
TR  & 0.0115 & 0.0113 & 0.0102 & \textbf{-0.0087} & 0.0408 \\
PPP  & 0.0106 & 0.0115 & 0.0113 & \textbf{-0.0061} & 0.0277 \\
MON  & 0.0052 & 0.0054 & 0.0037 & \textbf{-0.0060} & 0.0213\\
Oil  & -0.0054 & -0.0040 & -0.0050 & \textbf{-0.0078} & 0.0125 \\
Gold  & 0.0061 & 0.0061 & 0.0051 & \textbf{-0.0055} & 0.0377 \\
Copper  & -0.0063 & -0.0053 & -0.0039 & \textbf{-0.0058} & 0.0154 \\
\midrule
 \multicolumn{6}{l}{\textbf{h=3: 3-month ahead. }RW-SV: benchmark. RW-drift-SV: +0.0023} \\
IRP  & 0.0042 & 0.0062 & 0.0061 & \textbf{-0.0080} & 0.0247 \\
TR  & 0.0064 & 0.0093 & 0.0087 & \textbf{-0.0067} & 0.0351 \\
PPP  & 0.0025 & 0.0038 & 0.0025 & \boxed{\textbf{\textbf{-0.0108}}} & 0.0221 \\
MON  & 0.0043 & 0.0039 & 0.0027 & \textbf{-0.0083} & 0.0197 \\
Oil  & 0.0007 & 0.0013 & 0.0004 & \textbf{-0.0092} & 0.0258 \\
Gold  & 0.0071 & 0.0068 & 0.0069 & \textbf{-0.0022} & 0.0287 \\
Copper  & -0.0030 & -0.0027 & -0.0023 & \textbf{-0.0068} & 0.0200 \\
\midrule
\multicolumn{6}{l}{\textbf{h=6: 6-month ahead. }RW-SV: benchmark. RW-drift-SV: -0.0003} \\
IRP  & 0.0042 & 0.0043 & 0.0043 & \textbf{-0.0084} & 0.0430 \\
TR  & 0.0046 & 0.0029 & 0.0045 & \textbf{-0.0101} & 0.0375 \\
PPP  & -0.0004 & -0.0008 & -0.0002 & \textbf{-0.0096} & 0.0589 \\
MON  & 0.0015 & 0.0017 & 0.0008 & \textbf{-0.0098} & 0.0276\\
Oil  & -0.0041 & -0.0047 & -0.0033 & \boxed{\textbf{\textbf{-0.0107}}} & 0.0102 \\
Gold  & 0.0015 & 0.0015 & 0.0023 & \textbf{-0.0101} & 0.0318 \\
Copper  & 0.0016 & 0.0022 & 0.0022 & \textbf{-0.0082} & 0.0099 \\
\midrule
\multicolumn{6}{l}{\textbf{h=12: 12-month ahead. }RW-SV: benchmark. RW-drift-SV: -0.0024} \\
IRP  & 0.0021 & 0.0042 & 0.0033 & \textbf{-0.0095} & 0.0192 \\
TR  & 0.0027 & 0.0031 & 0.0024 & \boxed{\textbf{\textbf{-0.0144}}} & 0.0392 \\
PPP  & 0.0018 & 0.0016 & 0.0020 & \textbf{-0.0087} & 0.0285 \\
MON  & 0.0014 & 0.0009 & -0.0019 & \textbf{-0.0098} & 0.0280 \\
Oil  & 0.0008 & 0.0008 & -0.0004 & \textbf{-0.0101} & 0.0187 \\
Gold  & -0.0009 & -0.0012 & -0.0015 & \textbf{-0.0097} & 0.0155 \\
Copper  & 0.0046 & 0.0051 & 0.0049 & \textbf{-0.0063} & 0.0245 \\
 \bottomrule
\end{tabular}%

\vspace{0.2cm}
\footnotesize
\begin{tabular}{@{}p{\textwidth}@{}}
Note: CRPS values are averaged across all OOS periods and recorded as the distance to RW-SV without drift (benchmark values are 0.5226, 0.5246, 0.5264 and 0.5267 for each h). Bold numbers indicate the best-performing model in each row (Lower CRPS values indicate better predictive performance), and the boxed number indicates the best-performing model in each block. Model abbreviations: IRP = Interest Rate Parity, TR = Taylor Rule, PPP = Purchasing Power Parity, MON = Monetary Model.
\end{tabular}
\end{table}

\begin{table}[!htbp]
\centering
\caption{UK: Comparative Forecast Performance (1,3,6,12-period-ahead forecasting) Using Relative Root Mean Squared Forecast Error (RMSFE). Out-of-sample period: 2000/11 - 2017/6, in total 200 periods.}
\label{tab:ukrmsfe13612}

\begin{tabular}{lrrrrr} 
\toprule 
  & \multicolumn{3}{c}{Linear} & MSDSP & DSP \\
 \cmidrule(lr){2-4}\cmidrule(lr){5-5}\cmidrule(lr){6-6} 
\begin{tabular}[l]{@{}l@{}}Macro\\ Models\end{tabular}  & \begin{tabular}[c]{@{}c@{}}Expanding\\ Window\end{tabular} & \begin{tabular}[c]{@{}c@{}}100-rolling\\ Window\end{tabular} & \begin{tabular}[c]{@{}c@{}}50-rolling\\ Window\end{tabular}  &  &  \\
\midrule
\multicolumn{6}{l}{\textbf{h=1: 1-month ahead.} RW-SV: benchmark. RW-drift-SV: 1.0017} \\
IRP  & 1.0166 & 1.0167 & 1.0163 & \boxed{\textbf{\textbf{0.9841}}} & 1.0778 \\
TR  & 1.0311 & 1.0316 & 1.0310 & \textbf{0.9880} & 1.0687 \\
PPP  & 1.0161 & 1.0162 & 1.0163 & \textbf{0.9904} & 1.0419 \\
MON  & 1.0000 & 1.0002 & 1.0001 & \textbf{0.9891} & 1.0291 \\
Oil  & 0.9903 & 0.9899 & 0.9900 & \textbf{0.9870} & 1.0246 \\
Gold  & 1.0054 & 1.0052 & 1.0055 & \textbf{0.9911} & 1.0634 \\
Copper  & 0.9866 & 0.9866 & \textbf{0.9863} & 0.9898 & 1.0163\\
  \midrule
\multicolumn{6}{l}{\textbf{h=3: 3-month ahead. }RW-SV: benchmark. RW-drift-SV: 1.0049} \\
IRP  & 1.0152 & 1.0151 & 1.0151 & \textbf{0.9861} & 1.0427 \\
TR  & 1.0246 & 1.0248 & 1.0243 & \textbf{0.9879} & 1.0587 \\
PPP  & 1.0032 & 1.0032 & 1.0031 & \boxed{\textbf{\textbf{0.9743}}} & 1.0296 \\
MON  & 1.0016 & 1.0015 & 1.0016 & \textbf{0.9858} & 1.0260 \\
Oil  & 1.0009 & 1.0005 & 1.0005 & \textbf{0.9835} & 1.0454 \\
Gold  & 1.0107 & 1.0107 & 1.0105 & \textbf{0.9980} & 1.0488 \\
Copper  & 0.9954 & 0.9957 & 0.9952 & \textbf{0.9893} & 1.0308\\
\midrule
\multicolumn{6}{l}{\textbf{h=6: 6-month ahead. }RW-SV: benchmark. RW-drift-SV: 1.0002} \\
IRP  & 1.0086 & 1.0084 & 1.0081 & \textbf{0.9850} & 1.1073 \\
TR  & 1.0114 & 1.0117 & 1.0119 & \textbf{0.9841} & 1.0622 \\
PPP  & 0.9996 & 0.9996 & 0.9995 & \textbf{0.9841} & 1.1772 \\
MON  & 1.0010 & 1.0011 & 1.0009 & \textbf{0.9821} & 1.0465 \\
Oil  & 0.9919 & 0.9919 & 0.9919 & \boxed{\textbf{\textbf{0.9817}}} & 1.0017 \\
Gold  & 1.0028 & 1.0032 & 1.0031 & \textbf{0.9827} & 1.0637 \\
Copper  & 0.9976 & 0.9975 & 0.9974 & \textbf{0.9860} & 1.0083\\
\midrule
\multicolumn{6}{l}{\textbf{h=12: 12-month ahead. }RW-SV: benchmark. RW-drift-SV: 0.9961} \\
IRP  & 1.0034 & 1.0033 & 1.0034 & \textbf{0.9821} & 1.0270 \\
TR  & 1.0164 & 1.0164 & 1.0167 & \boxed{\textbf{\textbf{0.9687}}} & 1.0666 \\
PPP  & 1.0008 & 1.0010 & 1.0008 & \textbf{0.9843} & 1.0452 \\
MON  & 1.0000 & 1.0003 & 1.0001 & \textbf{0.9851} & 1.0413 \\
Oil  & 1.0017 & 1.0017 & 1.0018 & \textbf{0.9822} & 1.0307 \\
Gold  & 0.9906 & 0.9903 & 0.9906 & \textbf{0.9832} & 1.0168 \\
Copper  & 1.0057 & 1.0060 & 1.0059 & \textbf{0.9901} & 1.0387 \\
\bottomrule
 \end{tabular}

\vspace{0.2cm}
\footnotesize
\begin{tabular}{@{}p{\textwidth}@{}}
Note: RMSFE values are averaged across all OOS periods, and recorded as the ratio to RW-SV without drift (benchmark values are 0.9546, 0.9581, 0.9609 and 0.9609 for each h). Bold numbers indicate the best-performing model in each row (Lower RMSFE values indicate better predictive performance), and the boxed number indicates the best-performing model in each block. Model abbreviations: IRP = Interest Rate Parity, TR = Taylor Rule, PPP = Purchasing Power Parity, MON = Monetary Model.
\end{tabular}
\end{table}

\begin{table}[!htbp]
\centering
\caption{UK: Comparative Forecast Performance (1-period ahead forecasting) Using Coverage Rate across Categories. Out-of-sample period: 2000/11- 2017/6, in total 200 periods.}
\label{tab:ukcp02510}

\begin{tabular}{lrrrrr} 
\toprule 
  & \multicolumn{3}{c}{Linear} & MSDSP & DSP \\
 \cmidrule(lr){2-4}\cmidrule(lr){5-5}\cmidrule(lr){6-6} 
\begin{tabular}[l]{@{}l@{}}Macro\\ Models\end{tabular}  & \begin{tabular}[c]{@{}c@{}}Expanding\\ Window\end{tabular} & \begin{tabular}[c]{@{}c@{}}100-rolling\\ Window\end{tabular} & \begin{tabular}[c]{@{}c@{}}50-rolling\\ Window\end{tabular}  &  &  \\
\midrule
\multicolumn{6}{l}{\textbf{Top 2.5\%.}  RW-SV:1.5\%. RW-drift-SV: 1.5\%} \\
IRP  & 3.0\% & 3.0\% & 3.0\% & 2.5\% & 2.0\% \\
TR  & 3.5\% & 4.0\% & 3.5\% & 2.5\% & 2.0\% \\
PPP  & 3.0\% & 3.0\% & 2.5\% & 2.5\% & 2.5\% \\
MON  & 2.0\% & 2.0\% & 2.0\% & 3.0\% & 2.5\% \\
Oil  & 2.5\% & 2.0\% & 2.0\% & 2.5\% & 2.0\% \\
Gold  & 2.5\% & 2.5\% & 2.0\% & 2.5\% & 2.0\% \\
Copper  & 2.0\% & 2.5\% & 2.5\% & 2.5\% & 2.0\% \\
\textbf{Average} & 2.64\% & 2.71\% & 2.50\% & 2.57\% & 2.14\% \\ 
\midrule
\multicolumn{6}{l}{\textbf{Bottom 2.5\%.} RW-SV: 0.5\%. RW-drift-SV: 0.5\%} \\
IRP  & 1.0\% & 1.0\% & 1.0\% & 1.0\% & 1.0\% \\
TR  & 1.0\% & 1.0\% & 1.5\% & 1.0\% & 1.0\% \\
PPP  & 1.0\% & 1.0\% & 1.0\% & 1.0\% & 1.0\% \\
MON  & 0.5\% & 1.0\% & 1.0\% & 1.0\% & 1.0\% \\
Oil  & 0.5\% & 0.5\% & 1.0\% & 1.0\% & 1.0\% \\
Gold  & 0.5\% & 0.5\% & 1.0\% & 1.0\% & 0.5\% \\
Copper  & 1.5\% & 1.5\% & 1.0\% & 0.5\% & 1.0\% \\
\textbf{Average} & 0.86\% & 0.93\% & 1.07\% & 0.93\% & 0.93\% \\ 
\midrule\midrule
\multicolumn{6}{l}{\textbf{Top 10\%.}  RW-SV:4.5\%. RW-drift-SV: 5.5\%} \\
IRP  & 8.0\% & 7.0\% & 7.5\% & 8.0\% & 7.0\% \\
TR  & 8.0\% & 8.0\% & 8.0\% & 8.0\% & 6.5\% \\
PPP  & 8.0\% & 8.0\% & 8.5\% & 8.5\% & 8.5\% \\
MON  & 8.5\% & 8.5\% & 8.5\% & 8.0\% & 9.0\% \\
Oil  & 6.5\% & 6.5\% & 7.0\% & 8.0\% & 6.5\% \\
Gold  & 8.5\% & 8.5\% & 8.5\% & 8.0\% & 9.5\% \\
Copper  & 7.0\% & 7.5\% & 7.0\% & 8.0\% & 9.0\% \\
\textbf{Average} & 7.79\% & 7.71\% & 7.86\% & 8.07\% & 8.00\% \\
\midrule
\multicolumn{6}{l}{\textbf{Bottom 10\%.} RW-SV: 3.5\%. RW-drift-SV: 2.5\%} \\
IRP  & 7.0\% & 7.0\% & 7.5\% & 10.5\% & 10.0\% \\
TR  & 8.0\% & 8.5\% & 8.5\% & 10.5\% & 9.0\% \\
PPP  & 7.5\% & 6.5\% & 6.5\% & 11.0\% & 11.5\% \\
MON  & 7.0\% & 7.5\% & 7.5\% & 10.5\% & 10.0\% \\
Oil  & 9.5\% & 8.5\% & 8.5\% & 10.0\% & 9.0\% \\
Gold  & 6.5\% & 7.0\% & 7.0\% & 10.0\% & 9.0\% \\
Copper  & 8.5\% & 8.0\% & 8.0\% & 10.0\% & 8.0\% \\
\textbf{Average} & 7.71\% & 7.57\% & 7.71\% & 10.36\% & 9.36\% \\ 
\bottomrule
\end{tabular}%

\vspace{0.2cm}
\footnotesize
\begin{tabular}{@{}p{\textwidth}@{}}
Note: Coverage rate values are averaged across all OOS periods. Model abbreviations: IRP = Interest Rate Parity, TR = Taylor Rule, PPP = Purchasing Power Parity, MON = Monetary Model.
\end{tabular}
\end{table}

\begin{table}[]
    \centering
    \caption{UK: $95\%$ Model Confidence Set for 1,2-period-ahead forecasting}
    \label{tab:ukmcs_h12}
    \begin{tabular}{ll|rr|rr|rr}
\hline
\multirow{2}{*}{\begin{tabular}[c]{@{}l@{}}Estimation \\ Methods\end{tabular}} & \multicolumn{1}{c|}{\multirow{2}{*}{\begin{tabular}[c]{@{}l@{}}Macro \\ Models\end{tabular}}} & \multicolumn{2}{c|}{1. SFE} & \multicolumn{2}{c|}{2. Log Score} & \multicolumn{2}{c|}{3. CRPS} \\ \cline{3-8} 
 & \multicolumn{1}{c|}{} & \multicolumn{1}{r|}{h=1} & h=2 & \multicolumn{1}{r|}{h=1} & h=2 & \multicolumn{1}{r|}{h=1} & h=2 \\\hline
\multicolumn{2}{l|}{RW-SV} & \multicolumn{1}{r|}{14} & 11 & \multicolumn{1}{r|}{8} & 8 & \multicolumn{1}{r|}{14} & 9 \\
\multicolumn{2}{l|}{RW-drift-SV} & \multicolumn{1}{r|}{18} & 10 & \multicolumn{1}{r|}{9} & 9 & \multicolumn{1}{r|}{15} & 8 \\ \cline{1-2}
 & IRP & \multicolumn{1}{r|}{28} & 30 & \multicolumn{1}{r|}{-} & - & \multicolumn{1}{r|}{19} & 24 \\
 & TR & \multicolumn{1}{r|}{31} & 25 & \multicolumn{1}{r|}{-} & - & \multicolumn{1}{r|}{26} & 20 \\
\multirow{3}{*}{\begin{tabular}[c]{@{}l@{}}Linear \\ Expanding\\ Window\end{tabular}} & PPP & \multicolumn{1}{r|}{23} & 24 & \multicolumn{1}{r|}{-} & - & \multicolumn{1}{r|}{24} & 17 \\
 & MON & \multicolumn{1}{r|}{15} & 14 & \multicolumn{1}{r|}{-} & - & \multicolumn{1}{r|}{-} & - \\
 & Oil & \multicolumn{1}{r|}{12} & 21 & \multicolumn{1}{r|}{-} & - & \multicolumn{1}{r|}{9} & 21 \\
 & Gold & \multicolumn{1}{r|}{20} & 16 & \multicolumn{1}{r|}{-} & - & \multicolumn{1}{r|}{17} & 15 \\
 & Copper & \multicolumn{1}{r|}{10} & 7 & \multicolumn{1}{r|}{-} & - & \multicolumn{1}{r|}{10} & 12 \\\cline{1-2}
 & IRP & \multicolumn{1}{r|}{26} & 28 & \multicolumn{1}{r|}{22} & 23 & \multicolumn{1}{r|}{20} & 28 \\
 & TR & \multicolumn{1}{r|}{32} & 27 & \multicolumn{1}{r|}{24} & 20 & \multicolumn{1}{r|}{28} & 25 \\
\multirow{3}{*}{\begin{tabular}[c]{@{}l@{}}Linear \\ 100-rolling\\ Window\end{tabular}} & PPP & \multicolumn{1}{r|}{25} & 22 & \multicolumn{1}{r|}{20} & 12 & \multicolumn{1}{r|}{27} & 26 \\
 & MON & \multicolumn{1}{r|}{17} & 15 & \multicolumn{1}{r|}{18} & 18 & \multicolumn{1}{r|}{-} & - \\
 & Oil & \multicolumn{1}{r|}{13} & 20 & \multicolumn{1}{r|}{13} & 22 & \multicolumn{1}{r|}{11} & 22 \\
 & Gold & \multicolumn{1}{r|}{19} & 17 & \multicolumn{1}{r|}{15} & 15 & \multicolumn{1}{r|}{18} & 14 \\
 & Copper & \multicolumn{1}{r|}{8} & 8 & \multicolumn{1}{r|}{10} & 11 & \multicolumn{1}{r|}{12} & 7 \\\cline{1-2}
 & IRP & \multicolumn{1}{r|}{27} & 29 & \multicolumn{1}{r|}{21} & 24 & \multicolumn{1}{r|}{21} & 27 \\
 & TR & \multicolumn{1}{r|}{30} & 26 & \multicolumn{1}{r|}{23} & 19 & \multicolumn{1}{r|}{23} & 23 \\
\multirow{3}{*}{\begin{tabular}[c]{@{}l@{}}Linear \\ 50-rolling\\ Window\end{tabular}} & PPP & \multicolumn{1}{r|}{24} & 23 & \multicolumn{1}{r|}{19} & 14 & \multicolumn{1}{r|}{25} & 19 \\
 & MON & \multicolumn{1}{r|}{16} & 13 & \multicolumn{1}{r|}{17} & 17 & \multicolumn{1}{r|}{-} & 16 \\
 & Oil & \multicolumn{1}{r|}{11} & 19 & \multicolumn{1}{r|}{12} & 21 & \multicolumn{1}{r|}{8} & 18 \\
 & Gold & \multicolumn{1}{r|}{21} & 18 & \multicolumn{1}{r|}{16} & 16 & \multicolumn{1}{r|}{16} & 13 \\
 & Copper & \multicolumn{1}{r|}{9} & 6 & \multicolumn{1}{r|}{11} & 10 & \multicolumn{1}{r|}{13} & 10 \\\cline{1-2}
 & IRP & \multicolumn{1}{r|}{5} & 2 & \multicolumn{1}{r|}{6} & \textbf{1} & \multicolumn{1}{r|}{6} & 3 \\
 & TR & \multicolumn{1}{r|}{4} & 4 & \multicolumn{1}{r|}{4} & 2 & \multicolumn{1}{r|}{4} & 2 \\
 & PPP & \multicolumn{1}{r|}{7} & 12 & \multicolumn{1}{r|}{\textbf{1}} & 7 & \multicolumn{1}{r|}{7} & 11 \\
MSDSP & MON & \multicolumn{1}{r|}{6} & 9 & \multicolumn{1}{r|}{7} & 5 & \multicolumn{1}{r|}{5} & 6 \\
 & Oil & \multicolumn{1}{r|}{3} & 3 & \multicolumn{1}{r|}{6} & 6 & \multicolumn{1}{r|}{3} & 4 \\
 & Gold & \multicolumn{1}{r|}{2} & \textbf{1} & \multicolumn{1}{r|}{3} & 3 & \multicolumn{1}{r|}{\textbf{1}} & \textbf{1} \\
 & Copper & \multicolumn{1}{r|}{\textbf{1}} & 5 & \multicolumn{1}{r|}{2} & 4 & \multicolumn{1}{r|}{2} & 5 \\\cline{1-2}
 & IRP & \multicolumn{1}{r|}{34} & - & \multicolumn{1}{r|}{-} & - & \multicolumn{1}{r|}{31} & - \\
 & TR & \multicolumn{1}{r|}{-} & 32 & \multicolumn{1}{r|}{-} & 25 & \multicolumn{1}{r|}{-} & 29 \\
 & PPP & \multicolumn{1}{r|}{-} & - & \multicolumn{1}{r|}{-} & - & \multicolumn{1}{r|}{32} & - \\
DSP & MON & \multicolumn{1}{r|}{29} & 31 & \multicolumn{1}{r|}{-} & 13 & \multicolumn{1}{r|}{29} & 30 \\
 & Oil & \multicolumn{1}{r|}{-} & 34 & \multicolumn{1}{r|}{-} & - & \multicolumn{1}{r|}{-} & 32 \\
 & Gold & \multicolumn{1}{r|}{33} & 33 & \multicolumn{1}{r|}{-} & - & \multicolumn{1}{r|}{30} & 31 \\
 & Copper & \multicolumn{1}{r|}{22} & - & \multicolumn{1}{r|}{14} & - & \multicolumn{1}{r|}{22} & - \\ \hline
\end{tabular}%
 
  \vspace{0.2cm}
\footnotesize
\begin{tabular}{@{}p{\textwidth}@{}}
Note: The MCS results are based on the three loss functions considered. Rank M reflects a model's performance relative to others according to the selected loss function, with a lower rank indicating better performance (i.e., rank 1 corresponds to the best model). A dash (–) signifies that the model was eliminated in the procedure. Model abbreviations: IRP = Interest Rate Parity, TR = Taylor Rule, PPP = Purchasing Power Parity, MON = Monetary Model.
\end{tabular}
\end{table}
\begin{sidewaystable}[!htbp]
    \centering
    \caption{UK: Comparative Forecast Performance. Out-of-sample period: 2009/03 - 2017/6, in total 100 periods.}
    \label{tab:ukbps}
\resizebox{\textwidth}{!}{\begin{tabular}{lrrrrrrrrrrrr}
\toprule
\textbf{Assembly} &\textbf{h=1} & \textbf{h=2} & \textbf{h=3} & \textbf{h=4} & \textbf{h=5} & \textbf{h=6} & \textbf{h=7} & \textbf{h=8} & \textbf{h=9} & \textbf{h=10} & \textbf{h=11} & \textbf{h=12} \\ 
\toprule
\toprule
\textbf{LPL}\\
MSDSP &   \textbf{-142.3786} & \textbf{-144.8047} & \textbf{-142.3443} & \textbf{-144.7816} & \textbf{-141.3632} & \textbf{-141.8206} & \textbf{-143.1669} & \textbf{-142.9444} & \textbf{-144.5245} & \textbf{-142.6884} & \textbf{-142.1529} & \textbf{-142.9249} \\
DBPS &  -156.3835 & -162.8345 & -154.9650 & -153.9961 & -149.9553 & -148.0839 & -147.8373 & -144.2275 & -149.1719 & -147.4700 & {-142.3262} & -145.5031 \\
DSP &   -151.1876 & -158.5468 & -148.4923 & -151.4041 & -151.7629 & -152.0837 & -152.2112 & -146.9985 & -156.2246 & -149.0994 & -146.9559 & -150.3118 \\\hline
RW-SV &   -140.0417 & -139.8219 & -139.8489 & -139.8884 & -139.5621 & -139.9888 & -140.0941 & -140.2194 & -140.4038 & -140.1973 & -140.3494 & -140.3048 \\  
 \midrule
 \midrule
 \textbf{CRPS}\\
MSDSP  & \textbf{0.5692} & \textbf{0.5659} & \textbf{0.5500} & \textbf{0.5730} & \textbf{0.5523} & \textbf{0.5598} & \textbf{0.5716} & \textbf{0.5767} & \textbf{0.5638} & \textbf{0.5482} & \textbf{0.5506} & \textbf{0.5514} \\
DBPS  & 0.7306 & 0.7509 & 0.7048 & 0.7786 & 0.6978 & 0.6953 & 0.6938 & 0.6486 & 0.6909 & 0.6217 & 0.6626 & 0.6129 \\
DSP  & 0.6492 & 0.6853 & 0.6122 & 0.6364 & 0.6338 & 0.6345 & 0.6434 & 0.6320 & 0.6786 & 0.6034 & 0.5972 & 0.6178 \\\hline
RW-SV  & 0.5422 & 0.5427 & 0.5413 & 0.5404 & 0.5481 & 0.5481 & 0.5411 & 0.5496 & 0.5485 & 0.5572 & 0.5475 & 0.5438 \\ 
\midrule
\midrule
\textbf{RMSFE}\\
MSDSP  & \textbf{1.0041} & \textbf{1.0208} & \textbf{0.9894} & \textbf{1.0302} & \textbf{0.9860} & \textbf{0.9945} & \textbf{1.0182} & \textbf{1.0224} & \textbf{1.0104} & \textbf{0.9918} & \textbf{0.9872} & \textbf{0.9886} \\
DBPS   & 1.0928 & 1.1784 & 1.0936 & 1.1284 & 1.0411 & 1.0622 & 1.0753 & 1.0499 & 1.0244 & 1.0187 & 0.9895 & 1.0051 \\
DSP   & 1.1289 & 1.2143 & 1.1061 & 1.1277 & 1.1113 & 1.1017 & 1.1414 & 1.1466 & 1.2314 & 1.0810 & 1.0432 & 1.0879 \\ \hline
RW-SV   & 0.9841 & 0.9867 & 0.9824 & 0.9805 & 0.9962 & 0.9920 & 0.9826 & 0.9958 & 0.9902 & 1.0080 & 0.9917 & 0.9848
 \\ 
\bottomrule
\bottomrule
\end{tabular}}

    \vspace{0.2cm}
    \footnotesize
    \begin{tabular}{@{}p{\textwidth}@{}}
    Note: All LPL values are computed as the summation over all OOS periods. CRPS and RMSFE values are averaged across all OOS periods. Bold numbers indicate the best-performing model in each row (Higher LPDR values or Lower CRPS and RMSFE values indicate better predictive performance). Model abbreviations: IRP = Interest Rate Parity, TR = Taylor Rule, PPP = Purchasing Power Parity, MON = Monetary Model. These single models are the forecast results from the previous section using Linear models. 
    \end{tabular}
\end{sidewaystable}
\newpage
\bibliographystyle{apalike}  
\bibliography{ref.bib}  %

\end{document}